\documentclass[twocolumn]{aastex631}

\usepackage{multirow}
\usepackage{booktabs}
\usepackage{tablefootnote}
\usepackage{amsmath}
\usepackage{array}
\usepackage{tikz}
\usepackage{xcolor}
\usepackage{makecell}
\usepackage{bm}

\newcommand\Tstrut{\rule{0pt}{2.6ex}}

\begin{document}
\title{Multi-band ALMA Polarization Observations of BHB07-11 Reveal Aligned Dust Grains in Complex Spiral Arm Structures}

\author[0000-0001-7387-3898]{Austen Fourkas}
\affiliation{Department of Astronomy, University of Illinois, 1002 West Green St, Urbana, IL 61801, USA}

\author[0000-0002-4540-6587]{Leslie W. Looney}
\affiliation{Department of Astronomy, University of Illinois, 1002 West Green St, Urbana, IL 61801, USA}

\author[0000-0001-7233-4171]{Zhe-Yu Daniel Lin}
\affiliation{Earth and Planets Laboratory, Carnegie Institution for Science, 5241 Broad Branch Rd. NW, Washington, DC 20015, USA}

\author{Martin Radecki}
\affiliation{Department of Astronomy, University of Illinois, 1002 West Green St, Urbana, IL 61801, USA}

\author[0000-0002-7402-6487]{Zhi-Yun Li}
\affiliation{University of Virginia, 530 McCormick Rd., Charlottesville, Virginia 22904, USA}

\author[0000-0002-6195-0152]{John J. Tobin}
\affil{National Radio Astronomy Observatory, 520 Edgemont Rd., Charlottesville, VA 22903 USA} 

\author{Ian W. Stephens}
\affiliation{Department of Earth, Environment, and Physics, Worcester State University, Worcester, MA 01602, USA}

\author[0000-0001-5811-0454]{Manuel Fern\'{a}ndez-L\'{o}pez}
\affiliation{Instituto Argentino de Radioastronom{\'i}a, CCT-La Plata (CONICET), C.C.5, 1894, Villa Elisa, Argentina}
\affiliation{Facultad de Ciencias Astron\'omicas y Geof\'isicas, Universidad Nacional de La Plata, Paseo del Bosque S/N, B1900FWA La Plata, Argentina}

\author[0000-0002-8537-6669]{Haifeng Yang}
\affiliation{Institute for Astronomy, School of Physics, Zhejiang University, Hangzhou, 310027 Zhejiang, China}

\author[0000-0003-4022-4132]{Woojin Kwon}
\affiliation{Department of Earth Science Education, Seoul National University, 1 Gwanak-ro, Gwanak-gu, Seoul 08826, Republic of Korea}
\affiliation{SNU Astronomy Research Center, Seoul National University, 1 Gwanak-ro, Gwanak-gu, Seoul 08826, Republic of Korea}

\author[0000-0003-2118-4999]{Rachel Harrison}
\affiliation{Department of Physics and Astronomy, Rice University, 6100 Main Street -- MS 108, Houston, TX 77005, USA}
\begin{abstract}

\noindent Polarization mode observations from the Atacama Large Millimeter/submillimeter Array (ALMA) are powerful tools for studying the dust grain populations in circumstellar disks. Many sources exhibit polarization signatures consistent with aligned dust grains, yet the physical origin of this alignment remains uncertain. One such source is BHB07-11, a Class I protobinary object in the Pipe Nebula with complex spiral arm structures in its circumbinary disk. While magnetic fields are often invoked to explain grain alignment in the interstellar medium, the contrasting conditions in circumstellar disk environments demand further investigation into grain alignment mechanisms. To determine BHB07-11's dominant polarization mechanism, we leverage ALMA polarization mode dust continuum observations in Bands 3 $(\lambda=3.1\,\text{mm})$, 6 $(\lambda=1.3\,\text{mm})$, and 7 $(\lambda=0.87\,\text{mm})$, in combination with high-resolution dust continuum and spectral line observations in Band 6. Observed polarization vectors in each band are consistent with emission from aligned grains and follow the structure of the spiral arms as shown in the high-resolution observations. Given the relationship between the observed polarization vector orientation and the spiral arms, we find that the polarization morphology is most consistent with grains aligned through a relative velocity flow between gas and dust in the spiral arms, as envisioned in the recently developed badminton birdie-like alignment mechanism, rather than alignment with a magnetic field or other known alignment mechanisms.
\begin{center}
    \textbf{Accepted to ApJ}
\end{center}
\end{abstract}



\section{Introduction} 
\label{sec:Introduction}
Since the advent of the Atacama Large Millimeter/submillimeter Array (ALMA), polarization from thermally emitting dust grains has been observed at multiple wavelengths in a myriad of circumstellar disks across various evolutionary stages \citep[e.g.,][]{kataoka_submillimeter_2016, kataoka_evidence_2017, stephens_alma_2017, hull_alma_2018,Alves_2018, ohashi_two_2018,Sadavoy_VLA1623_2018, harrison_dust_2019, aso_multi-scale_2021, hull_polarization_2022, tang_polarization_2023, Stephens_Aligned_2023, lee_polarization_2024, lin_Panchromatic_2024, harrison_protoplanetary_2024, liu_dust_2024, Looney_L1448_2025, ohashi_observationally_2025}. ALMA polarization mode observations are a uniquely useful tool for characterizing the dust populations in circumstellar disks, as the mechanisms that induce disk-scale polarization are directly dependent on dust grain size, geometry, and orientation. Thus, identifying the mechanisms that cause dust grain polarization can lead to a more complete understanding of circumstellar dust populations, and provide critical insight into the complex processes of planet formation and stellar evolution. 

A combination of theoretical and observational work has shown that dust grain polarization in disks can generally be attributed to two different mechanisms: dust self-scattering \citep{kataoka_millimeter-wave_2015, yang_inclination-induced_2016} and alignment of non-spherical grains \citep[e.g.,][]{lazarian_radiative_2007, lazarian_subsonic_2007, andersson_interstellar_2015, tazaki_radiative_2017, hoang_Internal_2022, lin_Panchromatic_2024}. Additionally, polarization morphologies have been found to vary across observing wavelengths. Polarization from dust self-scattering has primarily been observed in short wavelength ALMA bands, while polarization consistent with aligned grains is prevalent in longer wavelength bands \citep[e.g.,][]{kataoka_evidence_2017, stephens_alma_2017, mori_modeling_2021}. For example, \cite{lin_Panchromatic_2024} presented interferometric observations of HL Tau between $\lambda=870\,\mu\text{m}-7.0\,\text{mm}$ and found that polarization smoothly transitions from scattering at shorter wavelengths to grain alignment at longer wavelengths. Such transitions can be explained by differences in optical depth \citep{lin_Thermal_2022a}, as optical depth is systematically higher in shorter wavelength ALMA Bands. In the optically thick regime, polarization from dust self-scattering dominates over emission from aligned grains, as photons emitted by a given dust grain are more likely to interact with other grains within the disk and experience either scattering or absorption. Conversely, in the optically thin regime, the photon--grain interaction probability is reduced, and observations thus primarily reveal photons emitted directly from grains in the disk (i.e, polarization from aligned grains) as opposed to those subject to scattering. However, such wavelength-dependent transitions are not always present in protostellar systems. For instance, \cite{harrison_protoplanetary_2024} found polarization from scattering in RY Tau and MWC 480 in both Bands 3 and 7. Given the complex nature of circumstellar environments and non-uniformity of polarization morphologies across ALMA bands, it is clear that multi-band observations are necessary to achieve a comprehensive understanding of scattering and grain alignment mechanisms.

While grain alignment through magnetic fields remains a useful explanation for observed disk-scale polarization morphologies, recent theoretical work has demonstrated that mechanical grain alignment facilitated by a relative gas--dust velocity flow can dominate over magnetic field alignment under the proper dense disk conditions \citep[The Badminton Birdie mechanism ---][]{lin_badminton_2024}. \cite{Looney_L1448_2025} applied this theory to the proto-multiple system L1448-IRS3B, which shows strikingly complex polarization morphology in ALMA Bands 4 and 7. They found that the polarization morphology in L1448-IRS3B's dense spiral arm structure is consistent with the Badminton Birdie mechanism, as polarization vectors in both bands are well aligned with the curvature of the arms, while magnetic field alignment is dominant in the outer regions of the system. These findings suggest that systems with assumed magnetic field grain alignment, particularly those with dense spiral arm structures, should be reinvestigated, including in the context of the Badminton Birdie grain alignment mechanism.

One such object is BHB07-11 (also called ``The Cosmic Pretzel"), a Class I protobinary system located in the Pipe Nebula at a distance of 163 pc \citep{dzib_distances_2018}. BHB07-11's constituent objects, BHB07-11A and BHB07-11B, have a combined mass of at most $2.25\pm0.13\,\,\text{M}_{\odot}$ and combined bolometric luminosity of $3.5\,\text{L}_{\odot}$ \citep{Alves_nature_2019}. \cite{Alves_2018} presented polarization mode dust continuum maps of BHB07-11 in ALMA Bands 3, 6, and 7, and inferred a magnetic field poloidal-to-toroidal strength ratio of 3:1 assuming grain alignment through the \textit{B}-RAT mechanism. Additionally, high-resolution ALMA Band 6 dust continuum observations presented in \cite{Alves_nature_2019} revealed BHB07-11's complex spiral arm structure. When overlaid on the high-resolution continuum image, polarization vectors in each ALMA band show remarkable alignment with the spiral arms, suggesting a direct relationship between the polarization and disk structure. In this work we investigate this relationship using recent theoretical and observational developments in the field of millimeter/submillimeter polarimetry \citep[e.g.,][]{yang_Does_2019, hoang_Internal_2022, Stephens_Aligned_2023, lin_badminton_2024, lin_Panchromatic_2024} to place new, more robust constraints on BHB07-11's polarization mechanism. To accomplish this task, we make use of the ALMA Band 3, 6, and 7 polarization mode observations initially presented in \cite{Alves_2018} and high-resolution dust continuum and spectral line observations from \cite{Alves_nature_2019}, which have not been combined in an analysis of the dust polarization to date.

Because relative velocity flow-induced alignment is directly tied to dust grain motion, BHB07-11 provides a unique opportunity to identify observational probes of dust grain kinematics, which, unlike line-of-sight gas kinematics, cannot be directly observed due to the thermal nature of dust grain emission. Further constraints on the polarization mechanism may aid in characterizing the disk-scale magnetic field proposed in \cite{Alves_2018}. To distinguish the mechanisms responsible for the observed polarization in the BHB07-11 system, we investigate polarization both from dust self-scattering and from aligned dust grains. First, we constrain contributions from self-scattering through the observed polarization percentages and vector orientations throughout the disk. Then, we leverage a series of morphological models and timescale calculations to identify possible contributions from various dust grain alignment mechanisms.

In Section \ref{sec:Observations} we present observational details and data reduction methods, and in Section \ref{sec:Observational Results} discuss the notable features of each observation. We analyze BHB07-11's polarization mechanism in Section \ref{sec:The Polarization Mechanism}, and discuss the implications of our analysis in Section \ref{sec:Discussion}. Finally, we present our conclusions in Section \ref{sec:Conclusions}.

\section{Observations} 
\label{sec:Observations}
\subsection{Continuum}
We make use of the ALMA Band 3, Band 6, and Band 7 dust continuum observations presented in \cite{Alves_2018}, which are centered at 97.5 GHz (3.1 mm), 233 GHz (1.3 mm), and 343.5 GHz (0.87 mm), respectively. The Band 3 and 7 observations were made during ALMA cycle-5 (project code: 2016.1.01186.S), and the Band 6 observations were made during ALMA cycle-2 (project code: 2013.1.00291.S). These observations were made in full polarization mode, which enabled the XX, YY, YX, and XY correlations for each antenna to be recorded. Information regarding the calibrators used for each observation block can be found in Section 2 of \cite{Alves_2018}. We also made use of high-resolution Band 6 dust continuum data centered at 225 GHz, which was recorded in the same cycle-5 observing program as the aforementioned Band 3 and Band 7 data. The Band 3, 6, and 7 polarization mode data were taken across 7 different execution blocks (EBs), and the high-resolution Band 6 data were taken in 3 EBs. Unless stated otherwise, any mention of Band 6 data refers to the cycle-2 observations. Brief observational details can be found in Table \ref{tab:Continuum details}.

\begin{table*}
\centering
    \caption{Continuum observation details}
    \label{tab:Continuum details}
    \begin{tabular}{cccccccccc}
    \toprule
    \multirow{2}{*}{Band} & \multirow{2}{*}{EBs} & \multirow{2}{*}{Date} & \multirow{2}{*}{Times} & \multirow{2}{*}{Antennas} & \multicolumn{4}{c}{Calibrators}\\
    \cmidrule(lr){6-9}
    &&\multirow{1}{*}{(y-m-d)}&\multirow{1}{*}{(UTC)}&\multirow{1}{*}{(12-m)}&Polarization & Flux & Phase & Bandpass\\
    \hline
    \multirow{2}{*}{Band 3} & \multirow{2}{*}{2} & \multirow{2}{*}{2017-11-14} & \multirow{1}{*}{16:38:39.8 -- 18:39:34.8} & \multirow{1}{*}{46} & \multirow{2}{*}{J1733-1304} & \multirow{2}{*}{J1733-1304} & \multirow{2}{*}{J1700-2610} & \multirow{2}{*}{J1924–2914} \\
    &&&\multirow{1}{*}{19:58:32.3 -- 21:32:29.4}& \multirow{1}{*}{44} &&&&&\\
    \hline
    \multirow{3}{*}{Band 6} & \multirow{3}{*}{3} & \multirow{2}{*}{2015-09-19} & \multirow{1}{*}{22:38:37.5 -- 00:21:02.4} & \multirow{1}{*}{35} & \multirow{3}{*}{J1751-0939} & \multirow{3}{*}{Titan} & \multirow{3}{*}{J1713–2658} & \multirow{3}{*}{J1924–2914} \\
    && \multirow{2}{*}{2015-09-20} &\multirow{1}{*}{00:27:27.3 -- 01:46:02.7}& \multirow{1}{*}{35} &&&&&\\
    &&&\multirow{1}{*}{02:06:04.7 -- 02:46:46.3} &\multirow{1}{*}{35}&&&&&\\
    \hline
    \multirow{3}{*}{Band 6 [HR]\tablenotemark{a}}& \multirow{3}{*}{3} & \multirow{1}{*}{2017-11-05} & 
    \multirow{1}{*}{20:57:27.9 -- 22:26:26.7} &\multirow{1}{*}{48}&\multirow{3}{*}{\nodata}&\multirow{3}{*}{J1733-1304}&\multirow{3}{*}{J1700-2610}&\multirow{2}{*}{J1924-2914}\\
    &&\multirow{1}{*}{2017-11-23}& \multirow{1}{*}{15:36:09.2 -- 17:04:13.1} &\multirow{1}{*}{43}&&&&&\\
    &&\multirow{1}{*}{2016-11-01}& \multirow{1}{*}{16:19:48.9 -- 17:15:19.5} &\multirow{1}{*}{40}&&&&\multirow{1}{*}{J1517-2422}\\
    \hline
    \multirow{2}{*}{Band 7} & \multirow{2}{*}{2} & \multirow{2}{*}{2017-05-11} & 
    \multirow{1}{*}{05:10:09.8 -- 07:04:55.8} &\multirow{1}{*}{45}&\multirow{2}{*}{J1733-1304} & \multirow{2}{*}{J1733-1304} & \multirow{2}{*}{J1700-2610} & \multirow{2}{*}{J1733-1304}\\
    &&& \multirow{1}{*}{07:06:56.6 -- 08:38:16.8} &\multirow{1}{*}{45}&&&&\\
    \bottomrule
    \end{tabular}
    \tablenotemark{a} High-resolution Band 6 data were not recorded in full polarization mode, which is why there is no polarization calibrator.
\end{table*}

Calibration for each band was performed using the \texttt{ScriptForPI} and \texttt{ScriptForPolCal} scripts provided by the ALMA team in combination with the CASA 5.1.0 ALMA pipeline \citep{TheCasaTeam_2022, Hunter_2023}. Prior to deconvolution, we performed self-calibration using the \texttt{auto\textunderscore selfcal}\footnote{\url{https://github.com/jjtobin/auto_selfcal}} \texttt{v1.1.0} script with CASA 6.5.5. To maintain a constant reference antenna throughout the self-calibration process, all instances of the \texttt{refant} parameter were changed from \texttt{flex} to \texttt{strict}. The Stokes \textit{I} results of the self-calibration process for each band are presented in Table \ref{tab:Self calibration}. In addition to phase calibration the \texttt{auto\textunderscore selfcal} script performs amplitude calibration if enough runs are successful. A successful run occurs when \texttt{auto\textunderscore selfcal} raises the signal to noise ratio (S/N) or dynamic range of the image, or the beam size does not significantly increase between successive solution intervals. Amplitude calibration was performed for the Band 6, high-resolution Band 6, and Band 7 data, but the limited S/N of the Band 3 data prevented the script from performing amplitude self-calibration. Nevertheless, phase self-calibration was performed for all three bands.

\begin{table*}
\centering
    \caption{Stokes \textit{I} self-calibration results. Values are truncated according to each ALMA Band's RMS accuracy.}
    \label{tab:Self calibration}
    \begin{tabular}{ccccccc}
    \toprule
        \multirow{3}{*}{Band} & \multicolumn{3}{c}{Initial} & \multicolumn{3}{c}{Final} \\
        \cmidrule(lr){2-4}\cmidrule(lr){5-7}
        &Flux&S/N&RMS&Flux&S/N&RMS\\
        &[mJy]&&[mJy beam$^{-1}$]&[mJy]&&[mJy beam$^{-1}$]\\

        \hline

        Band 3\tablenotemark{a} &19.290&182&0.009&21.540&184&0.008\Tstrut\\
        Band 6\tablenotemark{b} &488.80&229&0.10&713.90&680&0.04\\
        Band 6 [HR]\tablenotemark{c}&479.34&155&0.03&370.17&377&0.02\\
        Band 7\tablenotemark{d} &1159.3&184&0.3&1492.7&903&0.1\\

    \bottomrule
    \end{tabular}
    \tablenotetext{a}{Selfcal stopped due to low S/N. Final successful run: \texttt{solint=inf\textunderscore EB}.}
    \tablenotetext{b}{Selfcal finished all runs. Final successful run: \texttt{solint=inf\textunderscore ap}.}
    \tablenotetext{c}{Selfcal stopped due to low S/N. Final successful run: \texttt{solint=300s\textunderscore ap}.}
    \tablenotetext{d}{Selfcal finished all runs. Final successful run: \texttt{solint=inf\textunderscore ap}.}
    
\end{table*}
Imaging was performed in CASA 6.5.5 using the \texttt{tclean} task with \texttt{auto-multithresh} masking \citep{Kepley_2020, TheCasaTeam_2022}. The \texttt{Briggs} weighting scheme was used with a \texttt{Robust} parameter of 0.5 for both the polarization mode and high-resolution data. This weighting scheme allowed us to best match the images shown in Figures 7 and 8 of \cite{Alves_2018} while also resolving relevant emission in all three bands, which is necessary for proper comparison between the results shown in this work and those in \cite{Alves_2018}. Stokes \textit{I}, \textit{Q}, and \textit{U} maps were produced individually to prevent improper masking between images. After imaging was performed, the Band 6 data were regridded using \texttt{imregrid} to match the coordinate centers of the Band 3 and Band 7 data.

This work focuses only on linear polarization, and thus we create polarization intensity maps using 
\begin{align}
P_{\ell}=\sqrt{Q^{2}+U^{2}}. 
\end{align}
We also define the polarization intensity noise as 
\begin{align}
\sigma_{P}=\sqrt{(\sigma^{2}_{Q} + \sigma^{2}_{U})/2}, 
\end{align}
where $\sigma_{Q}$ and $\sigma_{U}$ are the RMS values of the Stokes \textit{Q} and \textit{U} maps, respectively. Additionally, we followed the approach taken in \cite{wardle_Linear_1974}, \cite{Vaillancourt_2006}, and \cite{Hull_Plambeck_2015} to debias the linear polarization maps for each band. For low S/N polarized emission $\left(P_{\ell}<3\sigma_{P}\right)$, we used the probability density function (PDF):
\begin{align}
    \text{PDF}(P|P_{\ell,\sigma_{P}}) = \frac{P}{\sigma_{P}^{2}}I_{0}\left(\frac{PP_{\ell}}{\sigma_{P}^2}\right)\exp\left[-\frac{(P_{\ell}+P^{2})}{2\sigma_{P}^2}\right].
\end{align}
This PDF takes the measured linear polarization, $P_{\ell}$, its noise, $\sigma_{P}$, and a zeroth-order modified Bessel function of the first kind, $I_{0}$, and produces the true linear polarization intensity, $P$. For high S/N polarized emission $\left(P_{\ell}\geq3\sigma_{P}\right)$, we make use of the approximation

\begin{align}
P=\sqrt{Q^{2}+U^{2}-\sigma_{P}^{2}}
\end{align}
to debias the measured linear polarization. Lastly, we calculate the polarization angle, $\chi$, for a given map pixel using 

\begin{align}
\chi = \frac{1}{2}\arctan{\left(\frac{U}{Q}\right)},
\end{align}
with the uncertainty in polarization angle given by 
\begin{align}
\sigma_{\chi}=\frac{1}{2}\frac{\sigma_{P}}{P}.
\end{align}

\begin{table*}
    \centering
    \caption{Stokes map properties. Values are truncated according to each ALMA Band's RMS accuracy.}
    \label{tab:Stokes properties}
    \begin{tabular}{ccccccc}
    \toprule
         \multirow{2}{*}{Band} & \multirow{2}{*}{Beam size} & \multirow{2}{*}{Beam PA} & \multicolumn{4}{c}{$\sigma$ [mJy beam$^{-1}$]} \\
         \cmidrule{4-7}
         &[arcsec]&[degrees]&I&Q&U&P \\
         \hline
         Band 3&0.15$\times$0.08&-49.12&0.009&0.009&0.009&0.009 \\ 
         Band 6&0.22$\times$0.12&-87.51&0.04&0.03&0.03&0.03 \\ 
         Band 6 [HR]\tablenotemark{a}&0.053$\times$0.038&-85.96&0.03&\nodata&\nodata&\nodata \\ 
         Band 7&0.18$\times$0.16&-73.62&0.2&0.02&0.02&0.02 \\ 
    \bottomrule
    \end{tabular}
    \tablenotetext{a}{Q, U, and P data were not available for the high-resolution Band 6 observations.}
\end{table*}
\subsection{Spectral Lines}
We also make use of high-resolution H$_{2}$CO $(3_{2,1}\,\mbox{--}\,2_{2,0})$ observations, which were recorded in the same program as the high-resolution Band 6 continuum data. As such, the same calibrators and antenna configurations were used. Data were recorded in two execution blocks, the first on 2017-11-05 between 20:57:27.9 and 22:26:26.7 UTC, and the second on 2017-11-23 between 15:36:09.2 and 17:04:13.1 UTC. 
As with the continuum data, calibration was performed using the provided \texttt{ScriptForPI} and the CASA 5.1.0 ALMA pipeline, while imaging was completed with CASA 6.5.5. Many of the same core cleaning parameters used for creating continuum maps were also used for creating line maps. The self-calibration solutions applied to the high-resolution Band 6 continuum data were not applied to these data. We produced a final data cube ranging from $v_{\text{LSR}}=-10$ to $v_{\text{LSR}}=+20$ km s$^{-1}$ with a channel width of 0.1 km s$^{-1}$, which encompasses relevant blue- and redshifted disk emission with respect to the systemic velocity of $3.6$ km s$^{-1}$. This cube has a beam size of $0.10\arcsec\times0.07\arcsec$, beam position angle of $82.0^{\circ}$, and RMS of 2 mJy beam$^{-1}$.

\section{Observational Results}
\label{sec:Observational Results}
\begin{figure*}
    \centering
    \includegraphics[width=\linewidth]{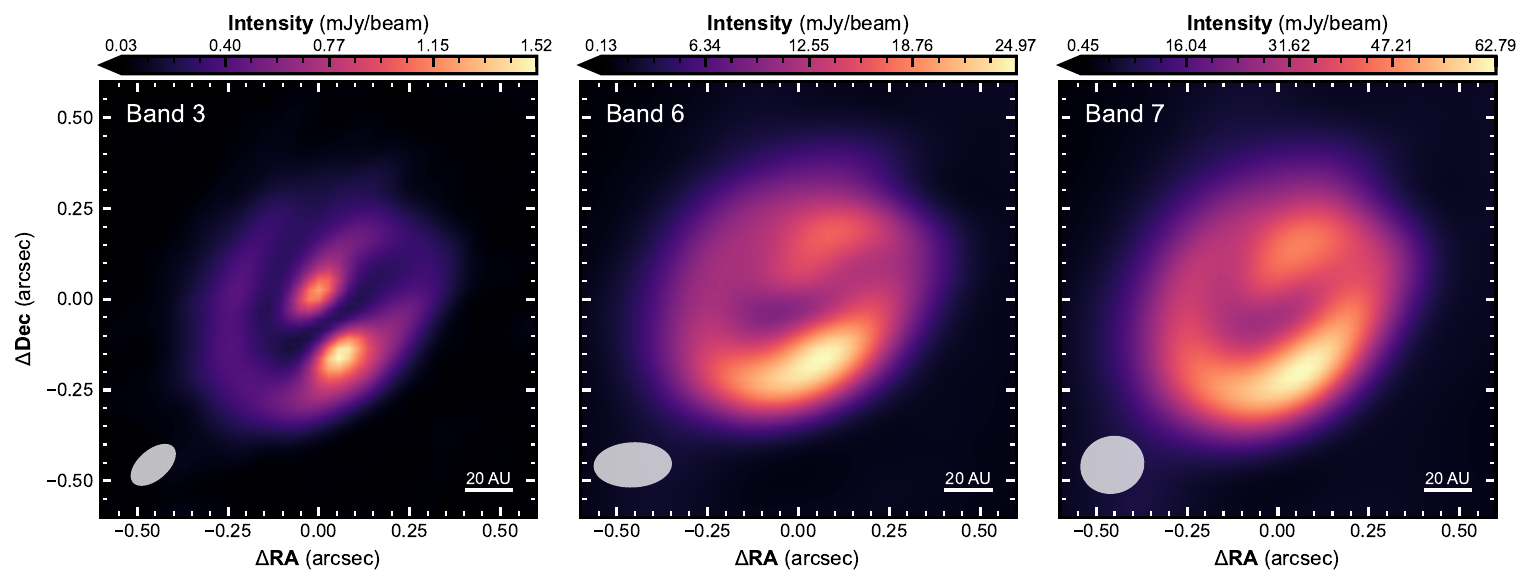}
    \caption{ALMA Band 3 (left), 6 (middle), and 7 (right) dust continuum maps of BHB07-11. Stokes \textit{I} emission above a 3$\sigma$ cutoff is shown in each panel. The beam size and position angle on the sky are shown in the bottom left corner of each image. All masked emission is shown in black as indicated by the colorbar arrow.}
    \label{fig:Band 367 I}
\end{figure*}

In Figure \ref{fig:Band 367 I} we show dust continuum (Stokes \textit{I}) maps of BHB07-11 taken in Bands 3, 6, and 7. BHB07-11's circumbinary disk is the primary target of the presented observations. The Band 3 dust continuum map highlights the spiral structure connecting BHB07-11A (northern intensity peak) and BHB07-11B (southern intensity peak). This structure is also apparent in the Band 6 and 7 continuum maps, although at lower resolution due to the shorter maximum baselines used in the observing configuration for the two bands. The locations of the intensity peaks also differ between bands. The Band 3 continuum map shows the BHB07-11A intensity peak at the end of the spiral structure, while the Band 6 and 7 maps lack a corresponding peak in the same location. These two maps instead show more intense emission along the spiral structure to the northwest of the BHB07-11A intensity peak. The BHB07-11B intensity peak, however, is present in the same location across all three bands. BHB07-11 is also host to a diffuse envelope which encompasses the circumbinary disk, and is present in each of the three bands (\citealp[see][]{Alves_2018} -- Figure 1).

\begin{figure}
    \centering
    \includegraphics[width=\linewidth]{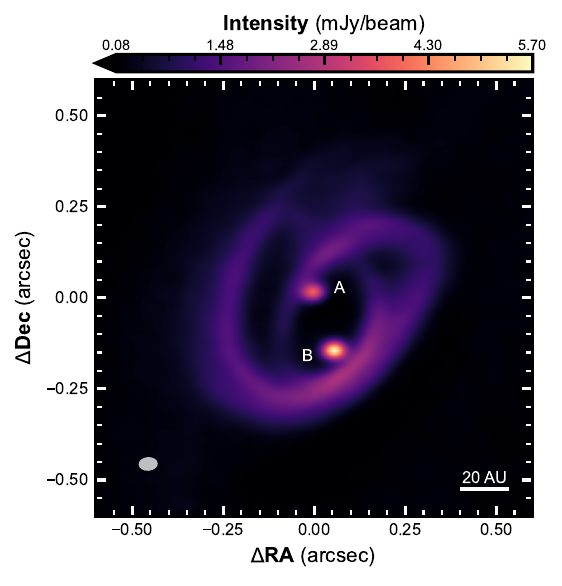}
    \caption{A high-resolution ALMA Band 6 dust continuum map of BHB07-11. BHB07-11A and BHB07-11B are labelled. Stokes \textit{I} emission is shown above a 3$\sigma$ cutoff. The beam size and position angle on the sky is shown in the bottom left corner of the map.}
    \label{fig:Band 6 HR}
\end{figure}

High-resolution Band 6 dust continuum images reveal well defined spiral arm structure throughout the disk (Fig. \ref{fig:Band 6 HR}). The spiral arms encircle the disk on its southern side, visually overlap towards the north, and reconnect in the center, forming the ``cosmic pretzel." In comparison to the low resolution continuum maps, a wealth of additional small-scale structure is resolved in the high-resolution map. In particular, a filament extending from BHB07-11A down to the southern portion of the spiral arms and a filament reaching from the northeast to the west of the disk are visible in the high-resolution dust continuum map. Additionally, extended low-intensity structure is visible on the northern edge of the disk, and is also seen in the low-resolution dust continuum images. The two emission peaks corresponding to BHB07-11A and BHB07-11B are also clearly defined. These peaks were interpreted as individual circumstellar disks around BHB07-11A and BHB07-11B in \cite{Alves_nature_2019}.

\begin{figure*}
    \includegraphics[width=\linewidth]{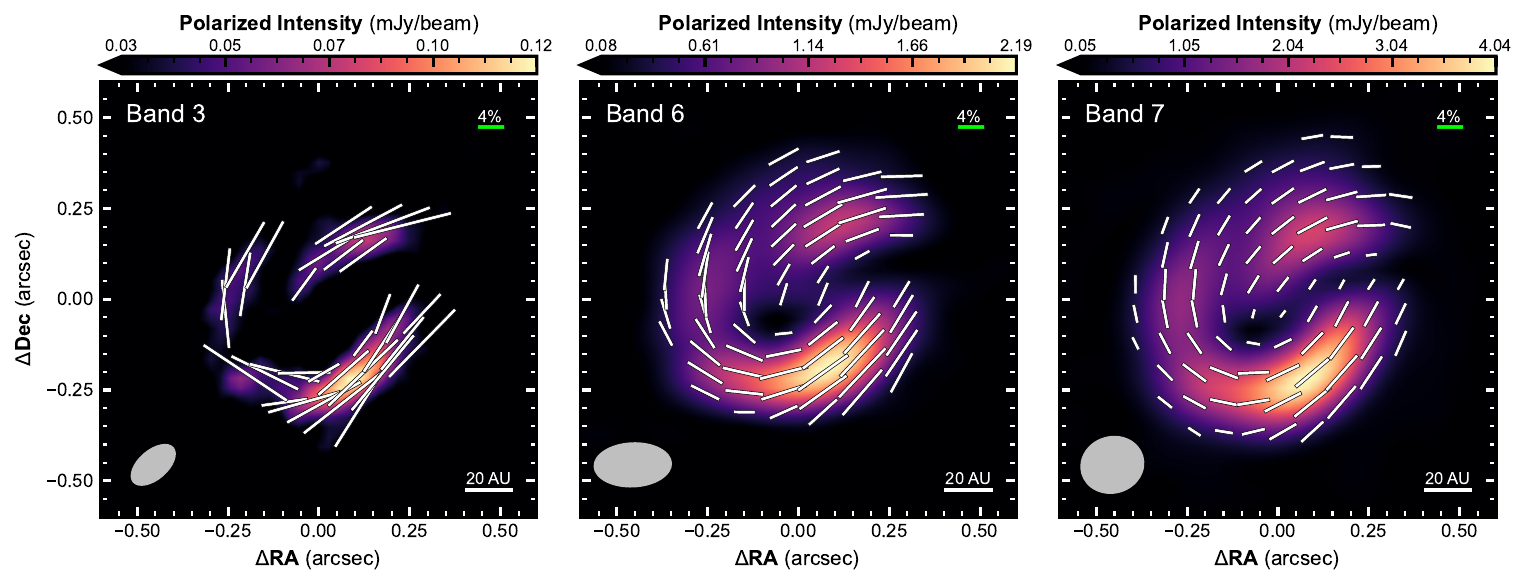}
    \caption{ALMA Band 3 (left), 6 (middle), and 7 (right) polarization mode dust continuum maps of BHB07-11. Polarized emission is shown above a 3$\sigma_{P}$ cutoff. Polarization vectors are shown in each frame, with their length proportional to the polarization percentage. Vectors are shown where polarized emission is above 4$\sigma_{P}$ in Band 3, 10$\sigma_{P}$ in Band 6, and 20$\sigma_{P}$ in Band 7, which limits the locations of the polarization vectors to lie within the disk. Vectors are Nyquist sampled across the beam major and minor axis in each map. A representative $4\%$ polarization vector key is shown in the top right of each frame, and the beam size and position angle on the sky are shown in the bottom left.}
    \label{fig:Band 367 P}
\end{figure*}

\begin{figure}
    \centering
    \includegraphics[width=\linewidth]{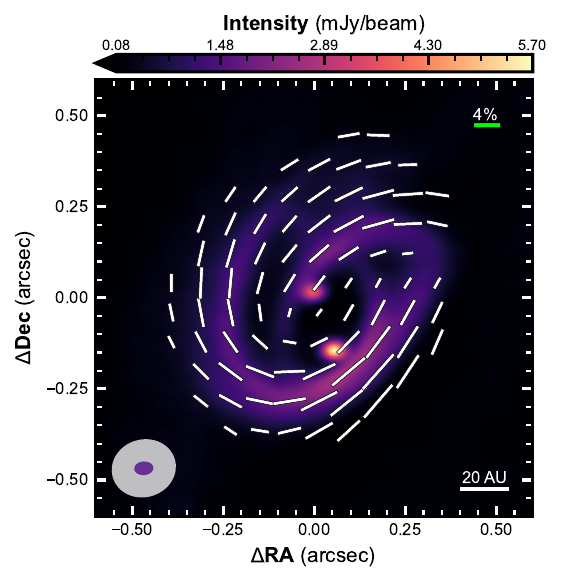}
    \caption{The high-resolution Band 6 dust continuum map shown in Fig. \ref{fig:Band 6 HR} with polarization vectors from the Band 7 image shown in Fig. \ref{fig:Band 367 P} overlaid. As with Figure \ref{fig:Band 367 P}, polarization vectors are cropped above 20$\sigma$ in polarized emission to restrict them to the disk. Vectors are shown at the same intervals as in Fig. \ref{fig:Band 367 P}. The Band 7 (white) and high-resolution Band 6 (purple) beam sizes and position angles on the sky are shown in the bottom left, and a green vector representing 4$\%$ polarization is shown in the top right.}
    \label{fig:Band 6 HR overlay}
\end{figure}

\begin{figure*}
    \centering
    \includegraphics[width=\linewidth]{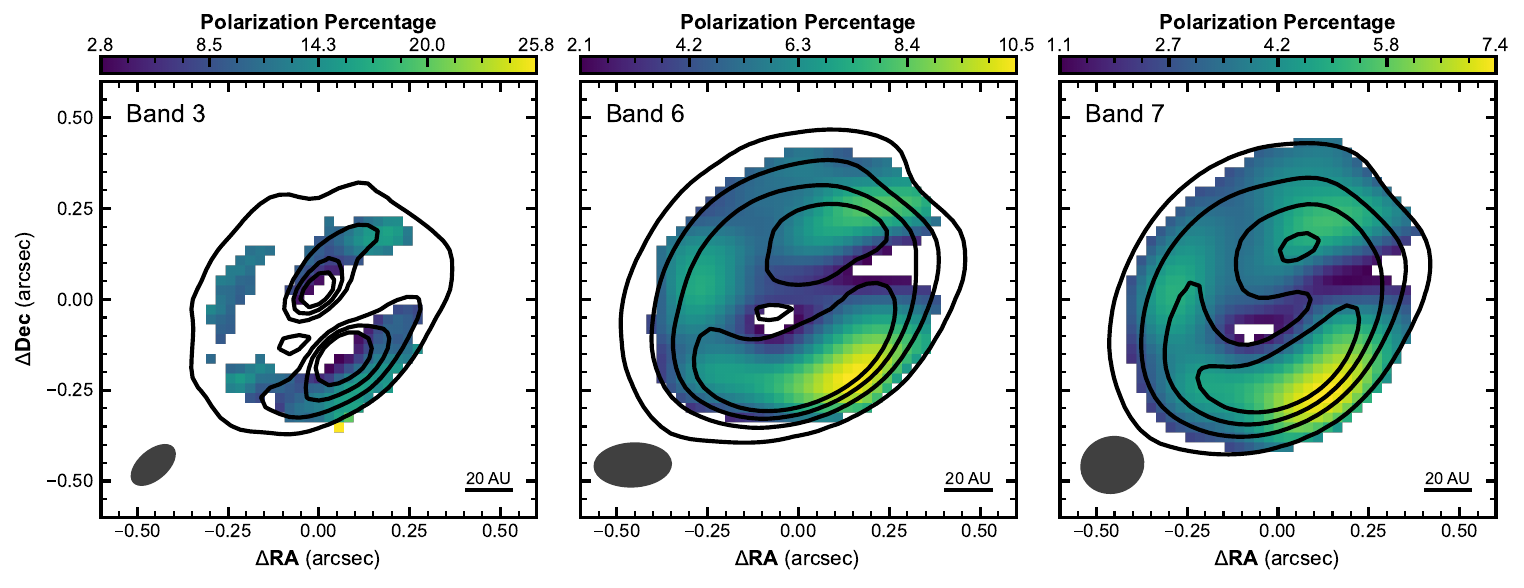}
    \caption{Polarization percentage maps constructed from the ALMA Band 3, 6, and 7 Stokes \textit{I} and polarized intensity images. The same polarized emission cuts from Figure \ref{fig:Band 367 P} are used to spatially restrict the polarization percentage. White pixels reflect masked data. The beam size and position angle are shown in the bottom left corner of each map.}
    \label{fig:Polarization fractions}
\end{figure*}

Polarization mode observations taken in Bands 3, 6, and 7 are shown in Figure \ref{fig:Band 367 P}. The polarized emission maps closely to the Stokes \textit{I} emission in each band, as the polarization morphology curls around the center of the disk and follows the spiral arms. The polarization vectors also show strong alignment with the spiral arm structure when overlaid on the high-resolution Band 6 data (Fig. \ref{fig:Band 6 HR overlay}). Polarized emission is absent in the westernmost portion of the disk, where the curvature in the Stokes \textit{I} maps is the highest. The emission peaks highlighted in the Stokes \textit{I} maps are also the most intensely polarized regions of the disk, although the continuum and polarization intensity peaks do not overlap exactly. Polarization intensity increases with increasing frequency; however, polarization fraction, defined as 
\begin{align}
\%p=\frac{P}{I},
\end{align}
decreases as frequency increases due to increased Stokes \textit{I} intensity in each band (Fig. \ref{fig:Polarization fractions}). This trend can additionally be attributed to optical depth differences across ALMA bands, as optical depth systematically increases with increasing frequency. As such, we find average beam sampled polarization fractions of $12.1\pm1\%$ for Band 3, $5.7\pm0.3\%$ for Band 6, and $4.0\pm0.2\%$ for Band 7.

\begin{figure*}
    \centering
    \includegraphics[width=\linewidth]{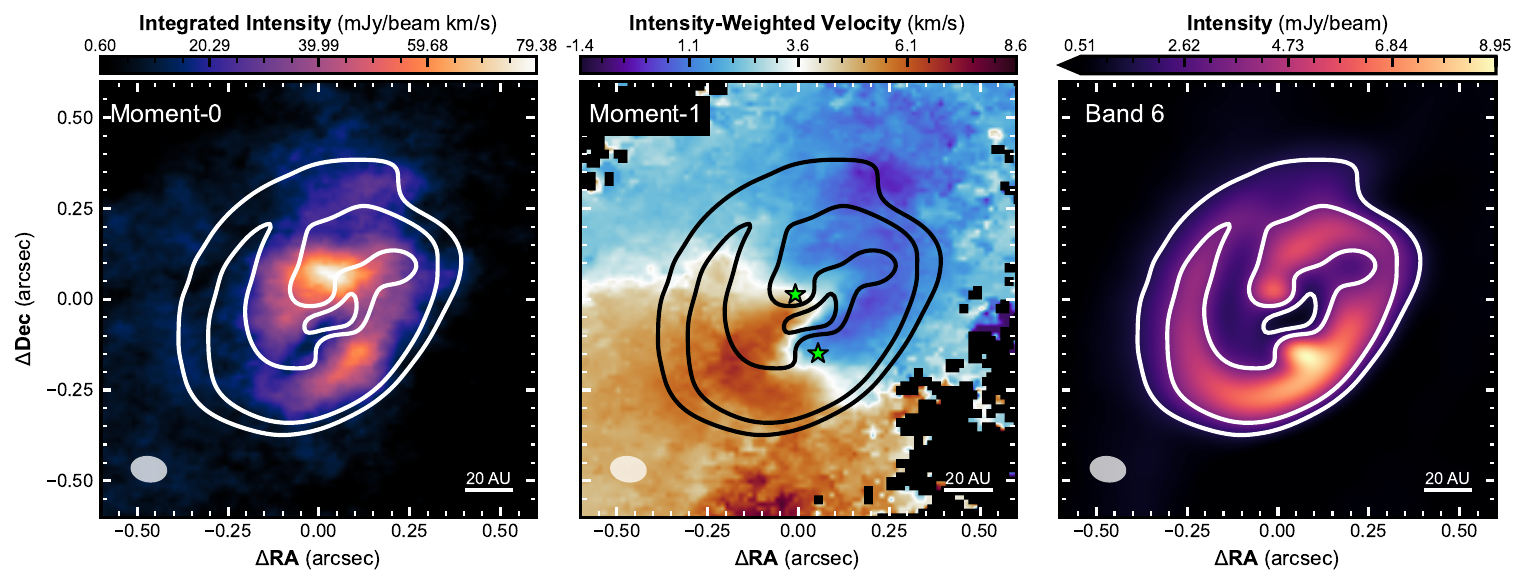}
    \caption{Moment-0 (left) and moment-1 (center) maps of H$_{2}$CO $(3_{2,1}\,\mbox{--}\,2_{2,0})$ integrated between $v_{\text{LSR}}=-1.4$ and $v_{\text{LSR}}=8.6$ km s$^{-1}$, and the high-resolution Band 6 dust continuum map (right). The Band 6 continuum is smoothed to the H$_{2}$CO map beam size of $0.10\times0.07$ arcseconds with a position angle of $82.0^{\circ}$. All maps only show emission above $3\sigma$. The overlaid contours represent the smoothed high-resolution Band 6 data at levels of 10$\sigma$ and 20$\sigma$ to highlight the shapess of the spiral arms. The beam size and position angle on the sky are shown in the bottom left of each panel.}
    \label{fig:Moment map}
\end{figure*}

Based on high-resolution $^{12}$CO $(\text{J}=2 \mbox{--} 1)$ observations, \cite{Alves_nature_2019} concluded that the spiral arms represent infalling gas from BHB07-11's large-scale circumbinary disk onto BHB07-11B, the southern protostellar object in the system. They also found a Toomre parameter of $Q\sim3.9$, indicating that the spiral arms are likely not results of disk fragmentation, but are rather accretion streamers feeding the two protobinary objects. Additionally, H$_{2}$CO $(3_{2,1}\,\mbox{--}\,2_{2,0})$ emission traces rotation on spatial scales similar to that of the spiral arms (Fig. \ref{fig:Moment map}), and at higher densities than CO \citep{mangum_formaldehyde_1993, tang_kinetic_2018}. Disk rotation is notably non-Keplerian, with BHB07-11A and BHB07-11B residing in cavities of redshifted and blueshifted gas, respectively. Gas velocities are highest surrounding BHB07-11B, with blueshifted emission peaking to the northwest and redshifted emission peaking to the east. This rotation profile may arise from perturbations to Keplerian rotation caused by the gravitational influence of the individual protobinary objects. The higher-velocity regions are not completely spatially coincident with the spiral arms, but do match the high-velocity $^{12}$CO emission reported in \cite{Alves_nature_2019}, although at lower overall velocities. Similarly, BHB07-11A straddles redshifted and blueshifted material, although at velocities closer to the systemic value than BHB07-11B. Rather than tracing accretion induced by the spiral arms, however, this emission may be indicative of the rotation of the large-scale circumbinary disk \citep{Alves_nature_2019}, as similar velocity structure extends well beyond the observed continuum emission.

\section{The Polarization Mechanism}
\label{sec:The Polarization Mechanism}

\subsection{Dust Self-Scattering}
To determine the dominant polarization mechanism within BHB07-11's disk, we first consider contributions from dust self-scattering. In the case of self-scattering, thermal emission from dust grains scatters off of other grains within the disk and can become polarized. Polarization from self-scattering can occur even if the incident photons are not polarized prior to scattering \citep{kataoka_millimeter-wave_2015}. When self-scattering occurs within the Rayleigh scattering regime, where the scattering grains are smaller than the wavelengths of the scattering photons, polarization is most efficient when the maximum grain size for solid, spherical grains in a given population is comparable to the observing wavelength $(a_{\text{max}}\sim\lambda/2\pi)$ \citep{kataoka_millimeter-wave_2015, kataoka_grain_2016}. Furthermore, photons that undergo scattering in the Rayleigh regime are maximally polarized when the angle of scattering between the incident and scattered photon is $\pi/2$ \citep{kataoka_millimeter-wave_2015}, and given that thermally emitting dust grains are primarily distributed throughout the disk midplane in moderately evolved disks, polarization vectors from self-scattering are characteristically aligned parallel to the disk minor axis in inclined disks \citep{yang_inclination-induced_2016}. Polarization from dust self-scattering produces polarization percentages in the range of $0.5\%-3\%$ in shorter wavelength ALMA Bands \citep[e.g.,][]{Sadavoy_VLA1623_2018, hull_alma_2018, ohashi_solving_2020, Stephens_Aligned_2023}.

As for BHB07-11, we find that self-scattering alone does not explain the observed polarization in any of the three ALMA bands examined here. While the previously mentioned characteristic vector morphology is applicable to axisymmetric or `disk-like' sources, the spiral arms fall into neither of these categories, meaning that scattering cannot be uniquely identified based solely on the observed polarization vector orientation in each band. Additionally, simulations from \cite{pohl_investigating_2016} showed that polarization vectors from scattering can be aligned azimuthally in face-on and inclined ringed disk structures. However, polarization from scattering can still be constrained based on the observed polarization percentages in each band.

\begin{figure*}
    \centering
    \includegraphics[width=\linewidth]{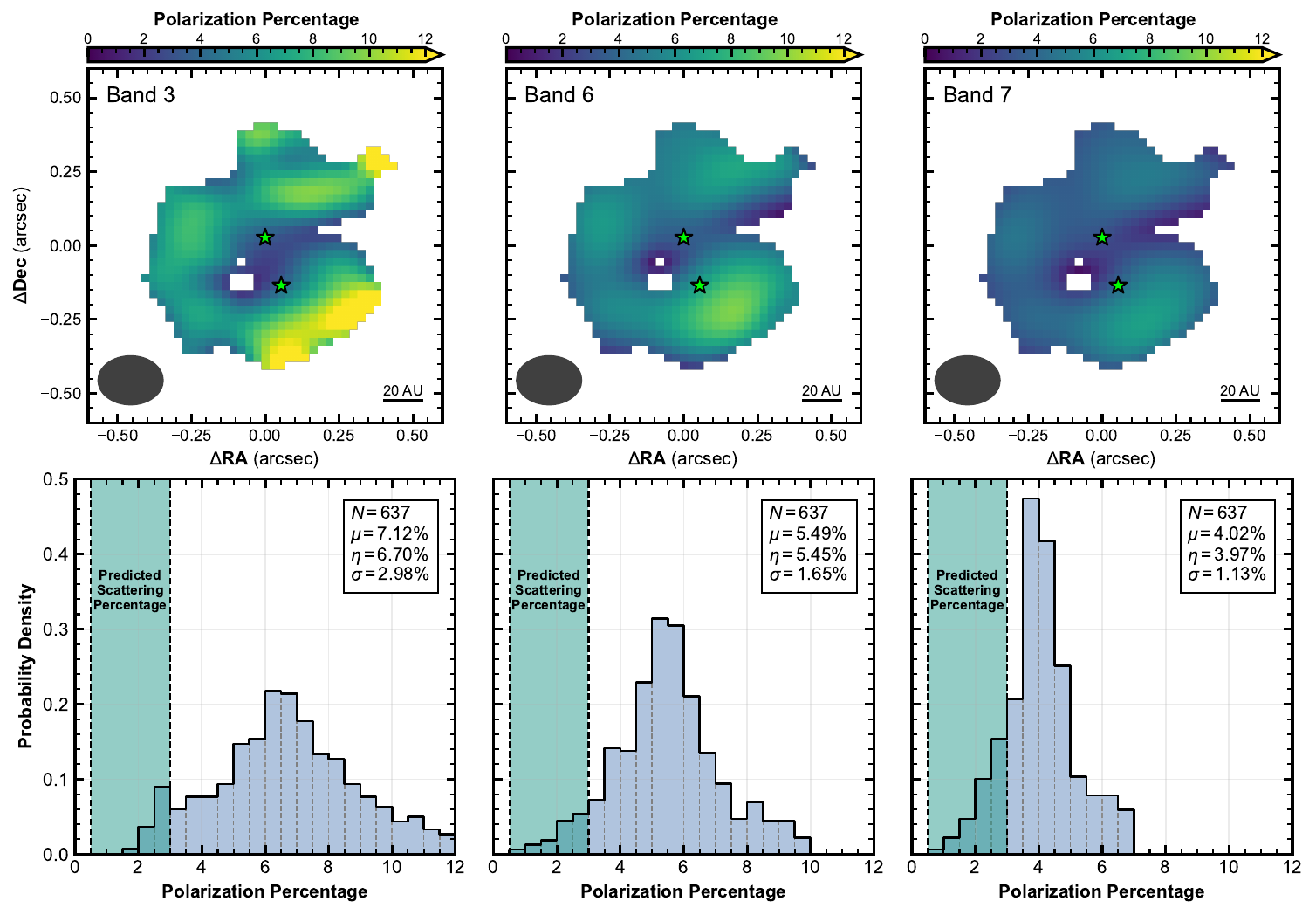}
    \caption{Top: Smoothed and \textit{uv}-limited polarization percentage maps for the Band 3, 6, and 7 observations used in this work. The upper and lower green stars represent the positions of BHB07-11A and BHB07-11B, respectively. The smoothing beam is shown in the bottom left of each frame, and has a size of $0.22\arcsec\times0.17\arcsec$ and a position angle $-90.0^{\circ}$. White pixels reflect masked data. Bottom: The binned polarization percentages for each map shown at the top of the figure. The statistics for each histogram are shown in the top right of each frame, and the predicted scattering percentage range is shaded. Bins have widths of $0.5\%$, which is larger than the maximum uncertainty in the polarization percentage in each band.}
    \label{fig:Percentage Histogram Combined}
\end{figure*}

To compare the polarization percentages in all three bands on a pixel-by-pixel basis, we smoothed all Stokes \textit{I}, \textit{Q}, and \textit{U} maps to a common beam with size $0.22\arcsec\times0.17\arcsec$ and position angle $-90.0^{\circ}$. We also restricted the \textit{uv} range of each map to $27-1310\,k\lambda$, which is the range of baselines shared by the polarization mode observations. Debiased polarization maps were then produced from the smoothed and \textit{uv} limited Stokes \textit{Q} and \textit{U} maps. We only considered pixels with Stokes \textit{I} and polarized emission above $3\sigma$ when creating the polarization percentage maps. Figure \ref{fig:Percentage Histogram Combined} shows the smoothed and \textit{uv} limited polarization percentage maps (top), and the binned distribution of polarization percentages for each band (bottom). We find that most pixels show polarization percentages outside the predicted range for scattering $(0.5\text{ -- }3\%)$, and that the peak polarization percentage increases as a function of wavelength. This trend does not match theoretical predictions for a combination of scattering and radiative alignment with grains of size $\sim150\,\mu\text{m}$ or smaller \citep{kataoka_evidence_2017}, which is larger than the predicted grain size in BHB07-11's spiral arms and envelope (see Section \ref{sec:Polarization Spectra}). 

Although scattering may make a contribution to the polarization in Bands 6 and 7, as these bands have systematically higher optical depth than Band 3, scattering alone cannot explain the observed polarization. Higher resolution ALMA polarization mode observations may reveal self-scattering on circumstellar disk scales $(\sim3\,\text{AU})$ around BHB07-11A and BHB07-11B, but current data indicate that polarization from dust self-scattering is not dominant in BHB07-11. These findings are consistent with those reported in \cite{Alves_2018}, and indicate that grain alignment must be the dominant source of polarization in BHB07-11.

\subsection{The Grain Alignment Mechanism}
\label{sec:The Grain Alignment Mechanism}
Given the negligible contributions from dust self-scattering, we next consider polarization from aligned grains. Grain alignment can be attributed to two general mechanisms: internal alignment and external alignment. Internal alignment concerns the alignment of a grain's axis of maximum moment of inertia with its angular momentum vector, $\bm{J}$, via energy dissipation mechanisms and relaxation effects \citep{hoang_Internal_2022}. External alignment is the alignment between $\bm{J}$ and an external field, including, but not limited to, a magnetic field, $\bm{B}$, a radiation field, $\bm{k}$, or a relative gas--dust velocity flow, $\bm{v}$ \citep{lazarian_radiative_2007, hoang_Internal_2022}.

Grains with helicity can become aligned through the application of radiative alignment torques (RATs) \citep{lazarian_radiative_2007}, which induce grain spin-up and enable alignment with various external fields. The RAT precession time represents the characteristic alignment timescale for RATs, and is instrumental in identifying the external fields with which dust grains become aligned. RAT-induced alignment with a magnetic field is known as the \textit{B}-RAT mechanism \citep{andersson_interstellar_2015}, and has often been invoked to explain disk-scale polarization from aligned grains \citep[e.g.,][]{Alves_2018, Sadavoy_VLA1623_2018, ohashi_two_2018, thang_evidence_2024, ohashi_observationally_2025}. \textit{B}-RAT alignment occurs when the Larmor precession time about the external magnetic field is shorter than the RAT precession time, and characteristically orients grains with their short axes parallel to the magnetic field direction when internally aligned. If the RAT precession time is instead shorter than the Larmor precession time, RATs can cause grain alignment with anisotropic radiation fields, which is known as the \textit{k}-RAT mechanism. Grain alignment through the \textit{k}-RAT mechanism produces circularly-oriented polarization vectors regardless of disk inclination \citep{yang_Does_2019}.

In addition to RATs, circumstellar dust grains can also undergo external alignment by mechanical alignment torques (MATs/METs), which are applied to grains via a relative velocity flow between the gas and dust in the disk \citep{lazarian_subsonic_2007}. Similar to RATs, METs can cause grain spin-up and enable alignment with external fields, and are characterized by the MET precession time. METs can induce grain alignment with magnetic fields if the Larmor precession time is shorter than the MET precession time, called the \textit{B}-MET mechanism. If MET precession dominates over Larmor precession then grains can become aligned with the MET-inducing relative gas--dust velocity flow, which is known as the \textit{v}-MET mechanism. Grains subject to the \textit{v}-MET mechanism will spin about their short axes, which are aligned parallel to the velocity flow.

Mechanical grain alignment can also occur without the influence of METs. The Gold mechanism \citep{gold_Alignment_1952} initially posits that supersonic relative gas--dust velocity flows can impart aligning torques on dust grains and produce linear polarization without the requirement of internal alignment. However, such velocities are not representative of gas drift velocities relative to dust in circumstellar disks, and the Gold mechanism cannot produce alignment in excess of $20\%$ \citep{Lazarian_Gold_1997}. On the other hand, \cite{lin_badminton_2024} showed that mechanical grain alignment can be achieved with subsonic gas drift velocities and yield high alignment percentages. In particular, they illustrated that non-spherical ellipsoidal dust grains (without helicity) can be aligned by drag-induced torques applied by a relative gas--dust velocity flow if an offset exists between the grain's geometric center and its center of mass. This alignment is similar to how a badminton birdie orients itself in the direction of its motion when traveling through the air, and is thus called the Badminton Birdie mechanism.

In the following analysis we consider three types of grain alignment fields and their respective mechanisms: a relative velocity flow between gas and dust, $\bm{v}$, a magnetic field, $\bm{B}$, and an anisotropic radiation field, $\bm{k}$. The corresponding grain alignment mechanisms are the Badminton Birdie and \textit{v}-MET mechanisms $(\bm{v})$, the \textit{B}-RAT and \textit{B}-MET mechanisms $(\bm{B})$, and the \textit{k}-RAT mechanism $(\bm{k})$. While circumstellar dust grains are likely triaxial and irregular, prolate and oblate ellipsoids can be used to describe the average, or effective, geometry of a population of grains \citep{lin_Panchromatic_2024}. Thus, for this analysis, we consider effectively prolate and oblate grain populations where applicable. Table \ref{tab:Timescale variables} lists all relevant variables and their uses in this analysis. 

We first constrain the orientation of the grain alignment field with respect to the spiral arms based on the observed polarization vectors. We then calculate a host of disk properties including dust grain size and optical depth, motivating calculations of the grain alignment timescale for magnetic field and velocity-flow induced grain alignment. Finally, we create a morphological model to identify polarization contributions from radiatively aligned dust grains.

\subsubsection{The Grain Alignment Field}
\begin{figure*}
    \includegraphics[width=\linewidth]{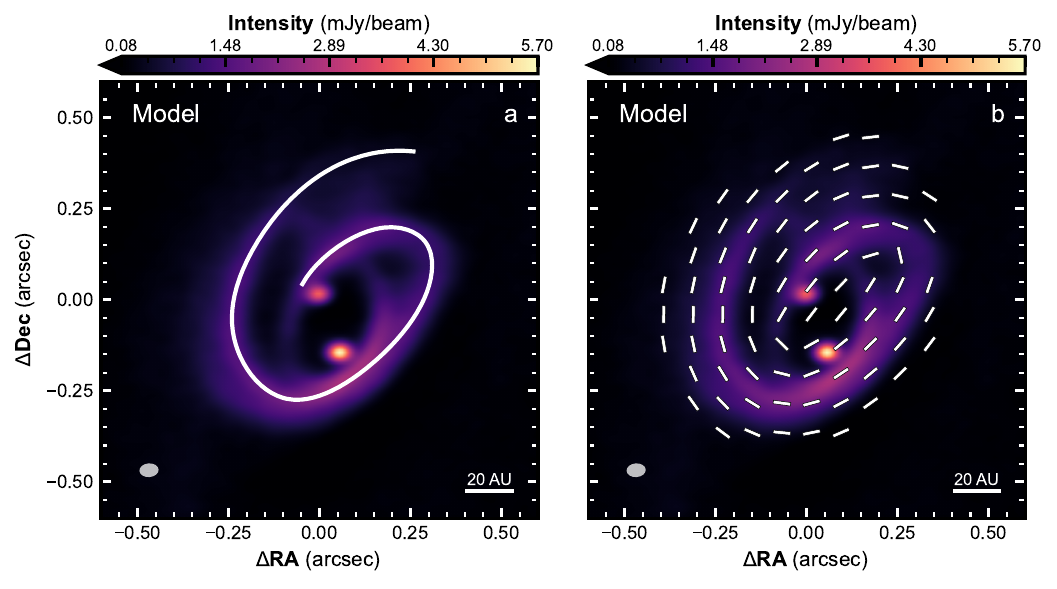}
    \centering
    \caption{Left: The high-resolution Band 6 dust continuum map from in Fig. \ref{fig:Band 6 HR} overlaid with the spline used to generate the spiral arm alignment model. The beam size and position angle on the sky are shown in the bottom left. Right: The high-resolution Band 6 dust continuum map from Fig. \ref{fig:Band 6 HR} overlaid with the polarization vectors from the spiral arm alignment model. Vectors are shown at the same intervals as in Fig. \ref{fig:Band 367 P} but are not scaled to the polarization percentage. The beam size and position angle on the sky are shown in the bottom left.}
    \label{fig:Splines}
\end{figure*}

Constraining the orientation of the grain alignment field is the first step in determining the mechanism responsible for the observed polarization in BHB07-11's circumbinary disk. Polarization vectors appear to trace the curvature of the spiral arms (Fig. \ref{fig:Band 6 HR overlay}), suggesting a relationship between the grain alignment field and the spiral arms. Motivated by L1448 IRS3B, which also showed polarization following the spiral arm morphology \citep{Looney_L1448_2025}, we create a morphological model to quantify the alignment between the observed polarization morphology in Band 7 and the spiral arm curvature. We choose to use the Band 7 observations for this comparison as they reveal the largest number of polarization vectors within the disk. To create the model, we first fit a spline to the shape of the spiral arms as seen in the high-resolution Band 6 continuum map (Fig. \ref{fig:Splines}a), and assigned a polarization angle to each pixel in the disk based on its proximity to the spline. The assigned angle is the slope of the spline at the closest point along the spline to the selected pixel. The spline itself is generated using the \texttt{radfil} package \citep{2018ApJ...864..152Z}, which requires a user-generated spatial region to constrain the spline fitting area. Because the spline is built using non-statistical processes, it is important to quantify how changes to the user defined region impact the spline fit. We first attributed a positional uncertainty of $\pm0.5$ pixels to each point on the region's edge, drew a value from a Gaussian distribution with $\mu=0$ and $\sigma=0.5$, and adjusted the point's position by the drawn value. This process was performed 1000 times (i.e., 1000 different regions are generated), and the spline averaged from each of the 1000 regions was used for the final model.

Because the Band 7 polarization map has a lower resolution, and thus a different pixel size, than the map used to create the polarization model, we rescaled the model to match the pixel size of the polarization mode observations. The selected pixel sizes for the polarization mode and high-resolution maps allow for nine high-resolution pixels (a $3\times3$ grid) to be averaged into a single low resolution pixel. The rescaled model vector morphology is shown in Figure \ref{fig:Splines}b. While necessary for proper comparison to the polarization mode observations, this method of rescaling leads to the loss of information at every pixel. To quantify this phenomenon, the standard deviation of each high-resolution $3\times3$ grid is used as the rescaling uncertainty in each respective low-resolution pixel. 

\begin{figure}
    \centering
    \includegraphics[width=\linewidth]{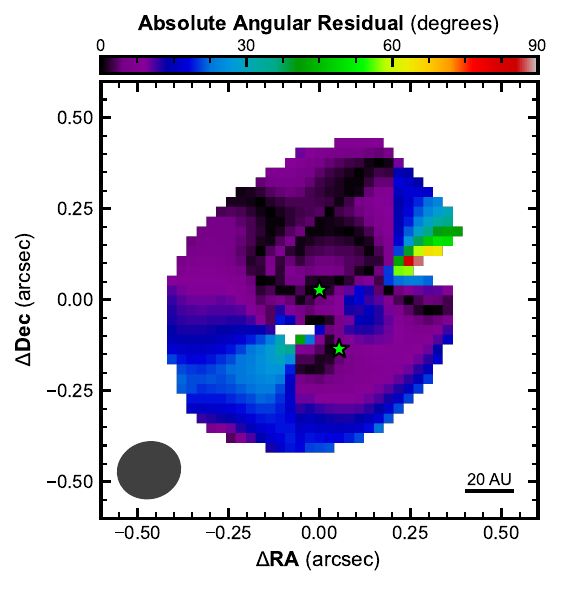}
    \caption{The spatial distribution of absolute angular residuals in the disk. The positions of BHB07-11A and BHB07-11B are represented by green stars, and the Band 7 beam size and position angle on the sky are shown in the bottom left.}
    \label{fig:difference map}
\end{figure}

After rescaling, we calculated the absolute minimum difference between the spline-based model and the observed Band 7 polarization morphology at each pixel in the disk, a method that accounts for the headless nature of the polarization vectors. The size of the Band 7 beam allowed us to create 30 independent beam sampled angular residual distributions, and we used Gaussian Kernel Density Estimation (KDE) to properly incorporate the model and observed polarization angle uncertainties into each residual distribution. The spatial distribution of absolute angular residuals throughout the disk is shown in Figure \ref{fig:difference map}. The residuals are lowest in the northeast and southwest portions of the disk, while two distinct high-residual regions are present in the southeast and northwest. The region in the southwest corresponds to polarization vectors seeming to `drift' around the curvature of the spiral arms in that area (see Fig. \ref{fig:Band 6 HR overlay}), which may be due to projection effects that limit our understanding of the three-dimensional geometry of the system. The high residuals in the northwest region are due to an abrupt change in the polarization on small angular scales and the limitations of the Band 7 beam size. The polarization flips from primarily vertically oriented to primarily horizontally oriented along the spiral arms (again, see Fig. \ref{fig:Band 6 HR overlay}), and the Band 7 beam size is too large to smoothly resolve this transition, thus resulting in poor agreement with the model generated using the high-resolution Band 6 beam size.

\begin{figure*}
    \includegraphics[width=\linewidth]{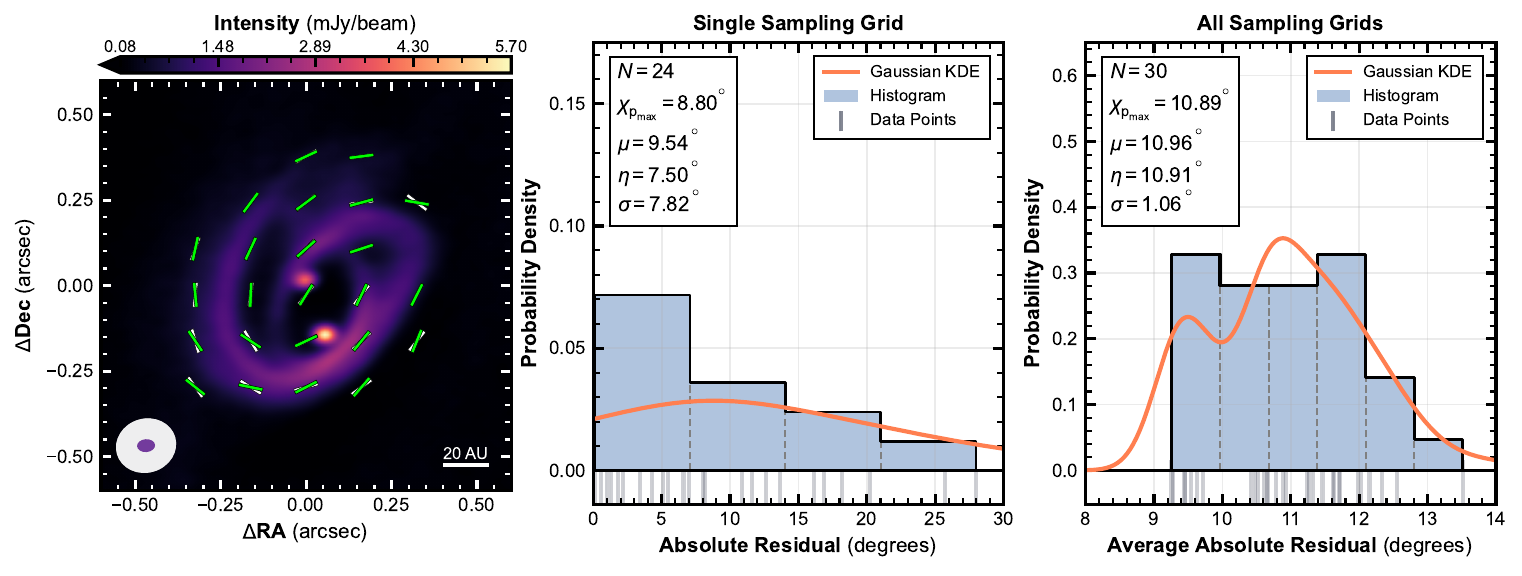}
    \caption{Left: The high-resolution Band 6 dust continuum map from Fig. \ref{fig:Band 6 HR} overlaid with the alignment model vectors (green) and observed Band 7 polarization vectors (white). Vectors are sampled once per Band 7 beam and are not scaled to the polarization percentage. The displayed vectors represent one of the 30 possible independent sampling grids. The Band 7 (white) and high-resolution Band 6 (purple) beam sizes and position angles on the sky are shown in the bottom left. Center: The distribution of absolute differences between the Badminton Birdie model and observed Band 7 polarization vectors shown in the left panel. The bandwidth of the KDE kernel is proportional to the uncertainty in the absolute difference between the model and observed polarization vectors. Right: The distribution of the average absolute residuals from each of the 30 independent sampling grids. The bandwidth of the KDE kernel is proportional to the uncertainty in each average absolute difference measurement. Statistics for the distributions shown in the middle and right panels are in the top left of each panel, respectively.}
    \label{fig:BB KDE plot}
\end{figure*}

A representative beam sampled residual distribution is shown in Figure \ref{fig:BB KDE plot}, along with the corresponding polarization vectors overlaid onto the spiral arms, as well as the distribution of the average absolute residual from each of the 30 individual distributions. The average absolute residuals, defined as
\begin{align}
|\Delta\chi|=|\chi_{\text{obs}}-\chi_{\text{model}}|,
\end{align}
for the spiral arm alignment model range from $|\Delta\chi|=9.3^{\circ}$ to $|\Delta\chi|=13.1^{\circ}$. In comparison, the best-fit magnetic field alignment model shown in \cite{Alves_2018} finds an average angular residual of $|\Delta\chi|=23.4^{\circ}$. Thus, we posit that the grain alignment field is strongly related BHB07-11's spiral arms, given the high degree of alignment between the observed and model polarization morphologies. This interpretation of the grain alignment field is consistent with that of \cite{Looney_L1448_2025}, which found that the spiral arms visible in L1448 IRS3B are also connected to the observed aligned polarization vectors. Given this interpretation, we can now constrain the grain alignment mechanism through the alignment field orientation.

\subsubsection{Polarization Spectra}
\label{sec:Polarization Spectra}
The requirement for grain alignment in multiple ALMA Bands allows for various disk parameters to be calculated which are useful in determining the grain alignment mechanism. Assuming the disk is geometrically thin, the combination of Stokes \textit{I} and polarization observations allow for optical depth, polarization in the optically thin limit, and dust temperature to be estimated using the techniques detailed in \cite{Hildebrand_Davidson_Dotson_Dowell_Novak_Vaillancourt_2000}. Specifically, the observed polarization percentage, $\%p_{\nu}$, and the intensity, $I_{\nu}$, at frequency $\nu$ from an isothermal, homogeneous slab of aligned grains are given by:
\begin{align} 
    \%p_{\nu}&=\frac{e^{-\tau}\sinh(P_{0}\tau_{\nu})}{1-e^{-\tau}\cosh(P_{0}\tau_{\nu})}\label{eq:Hildebrand 1}
\end{align}
and
\begin{align}
    I_{\nu}&=B_{\nu}(T)(1-e^{-\tau_{\nu}})\label{eq:Hildebrand 2}
\end{align}
respectively, where $\tau_{\nu}$ is optical depth, $B_{\nu}(T)$ is the Planck function at temperature $T$, and $P_{0}$ is the polarization in the optically thin limit. 

Using the observed values of $I_{\nu}$ and $\%p_{\nu}$ in Bands 3 and 7, which allows for the greatest frequency coverage of the observations included in this work, we iteratively solved equations \ref{eq:Hildebrand 1} and \ref{eq:Hildebrand 2} using the SciPy function \texttt{fsolve} for $P_{0}$, $T$, $\tau_{3}$, and $\tau_{7}$, where $\tau_{3}$ and $\tau_{7}$ are the optical depths in Bands 3 and 7, respectively. If \texttt{fsolve} failed to find a solution, the solution vector was all zeros, or any one solution was negative for a given pixel, the fit was rejected and the pixel was masked in the final map. To properly compare the two images we limited their \textit{uv} ranges to $27-1310\,k\lambda$ and smoothed them to a common beam of $0.185\arcsec\times0.165\arcsec$ with a position angle of $-70.0^{\circ}$. 

\begin{figure*}
    \centering
    \includegraphics[width=\linewidth]{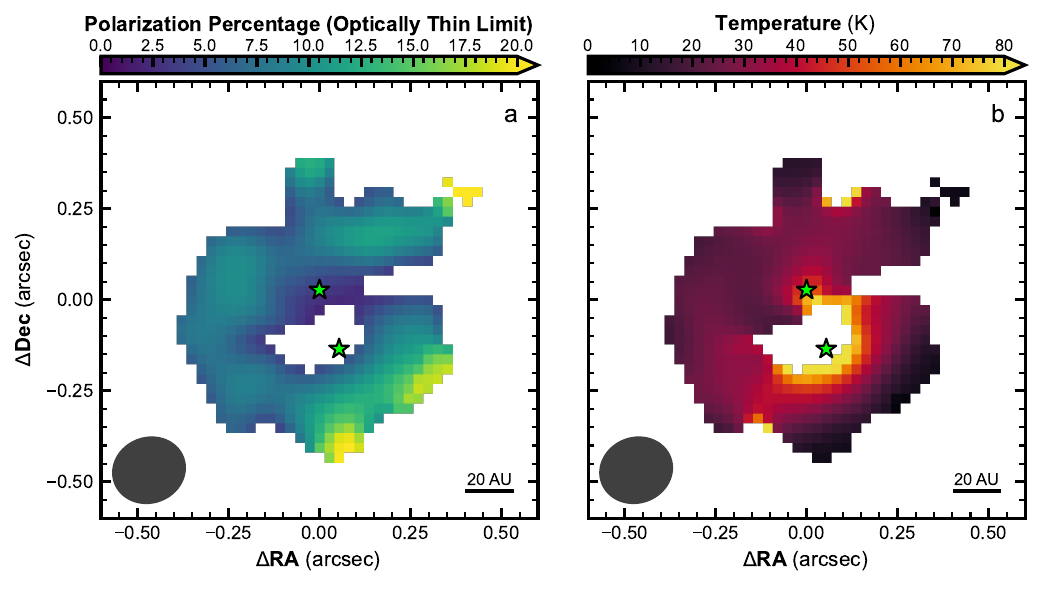}
    \includegraphics[width=\linewidth]{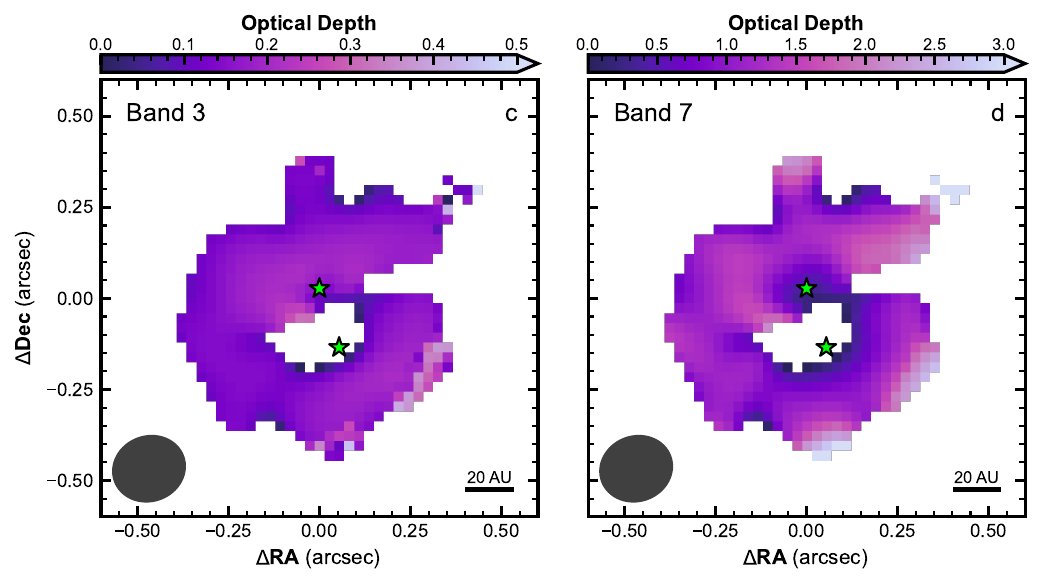}
    \caption{Estimates for (a) the intrinsic polarization, (b) the dust temperature, and (c) and (d) the optical depths in Band 3 and Band 7 using equations from \cite{Hildebrand_Davidson_Dotson_Dowell_Novak_Vaillancourt_2000}. Calculations were performed using Band 3 and Band 7 Stokes \textit{I} and \textit{P} data. Band 3 data were smoothed to the Band 7 beam. All Stokes \textit{I} and \textit{P} emission was cropped above 3$\sigma$ and 3$\sigma_{P}$, respectively. The positions of BHB07-11A and BHB07-11B are represented by green stars. The fit images do not directly match to the presented observations due to the pixel masking process, which is why the location of BHB07-11B is empty in each of the plots. The smoothed beam size and position angle on the sky are shown in the bottom left of each panel.}
    \label{fig:Fitting results}
\end{figure*}

Figure \ref{fig:Fitting results} shows the inferred maps of $P_{0}$, $T$, $\tau_{3}$, and $\tau_{7}$. We find values of $P_{0}$ primarily between $5-15\%$, which is in line with that of HL Tau \citep{Stephens_Aligned_2023, lin_Panchromatic_2024}. Furthermore, $\tau_{7}$ is systematically higher than $\tau_{3}$, which is expected due to increased dust opacity at shorter observing wavelengths. Using the size of the smoothed beam, we find average beam sampled values of $8.6\pm0.8\%$ for optically thin polarization percentage, $29.3\pm3.1\,\text{K}$ for dust temperature, and $0.16\pm0.013$ $\&$ $1.20\pm0.13$ for optical depth in Bands 3 and 7, respectively.

The Band 3 and 7 optical depth maps also allow for the spatial distribution of the dust opacity spectral index, $\beta$, to be found. The spectral index, $\alpha$, is given by:
\begin{align}
\alpha=\frac{\partial\log\left(S_{\nu}(\nu)\right)}{\partial\log\left(\nu\right)},
\end{align}
where $S_{\nu}(\nu)$ is the flux density at frequency $\nu$, and allows for the dust opacity spectral index to be estimated via 
\begin{align}
\label{eq:beta approx}
\beta \approx \alpha-2,
\end{align}
in the optically thin limit. When calculated in this way, optical depth can have significant impacts on the calculation of the spectral index, and thus mask the true value of the $\beta$. The dust opacity, $\kappa_{\nu}$, is related to the dust opacity spectral index through 
\begin{align}
\kappa_{\nu}\propto\nu^{\beta}, 
\end{align}
and because the optical depth is related to the opacity through 
\begin{align}
\tau_{\nu}=\rho_{d}\kappa_{\nu},
\end{align}
where $\rho_{d}$ is the dust density and is constant at both wavelengths, the dust opacity spectral index can be found with optical depth:
\begin{align}
    \beta = \frac{\log\left(\tau_{7}/\tau_{3}\right)}{\log\left(\nu_{7}/\nu_{3}\right)},
\end{align}
where $\nu_{3}$ and $\nu_{7}$ are the Band 3 and 7 observing frequencies. This technique provides a more robust method for calculating $\beta$ and removes the optical depth effects that lead to uncertainties in its measurement through Equation \ref{eq:beta approx}.

\begin{figure}
    \centering
    \includegraphics[width=\linewidth]{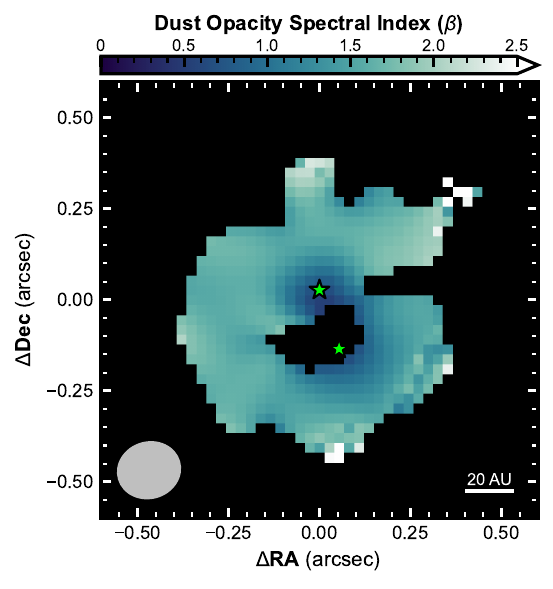}
    \caption{The spatial distribution of the dust opacity spectral index $(\beta)$ found using the optical depth maps from in Figs. \ref{fig:Fitting results}c and \ref{fig:Fitting results}d. The positions of BHB07-11A and BHB07-11B are represented by green stars, and the smoothed beam size and position angle on the sky are shown in the bottom left.}
    \label{fig:Beta map}
\end{figure}

The map of $\beta$ is shown in Figure \ref{fig:Beta map}. We find an average beam sampled dust opacity spectral index of $\beta=1.53 \pm0.07$, which is shallower than that of the interstellar medium (ISM) \citep[$\beta_{\text{ISM}}\approx1.7$,][]{Weingartner_and_Draine_ISM}. This result indicates that BHB07-11's grains have grown from the typical ISM size of $\sim0.1 \text{--} 1\,\mu\text{m}$ to $\sim10 \text{--} 50\,\mu\text{m}$ \citep{ysard_grains_2019, cacciapuoti_faust_2025}. The measured average value of $\beta$ in this work does not match the $\beta\approx1$ measurement reported in \cite{Alves_2018}, which suggests that BHB07-11's grains have experienced growth of up to millimeters in size. However, grains in BHB07-11's large-scale envelope ($\sim500\,\text{AU}$) were found to have a dust opacity spectral index of $1.50\pm0.08$ \citep{cacciapuoti_faust_2025}, which is consistent with the findings of this work. \cite{cacciapuoti_faust_2025} also found that grains in BHB07-11's envelope are smaller than the $\sim100\,\mu\text{m}$ predicted in previous works, as the calculated dust opacity spectral index is within a $1\sigma$ deviation from ISM values. Thus, we conclude that BHB07-11's grains on disk and envelope scales have experienced only slight growth in comparison to grains in the ISM, and are predicted to have sizes on the order of $\sim10 \text{--} 50\,\mu\text{m}$.

Self-scattering also places constraints on the maximum grain size $(a_{\text{max}})$ in the disk. In ALMA Band 7, where scattering is anticipated to dominate due to higher optical depth, scattering efficiency is highest when grains have a maximum size of $a_{\text{max}}\sim150\,\mu\text{m}$ \citep{kataoka_millimeter-wave_2015}. Given the average beam sampled optical depth in Band 7 ($\bar{\tau_{7}}=1.20\pm0.13$), we expect scattering to be present in the observed polarization if grains are of the appropriate size, as the observed emission is optically thick. We find  that scattering is not the dominant source of polarization in this band, further suggesting that the maximum grain size must be smaller than $150\,\mu\text{m}$, and is indeed closer to $\sim10\text{ - }50\,\mu\text{m}$.

\subsubsection{Alignment Timescales}
\label{sec:Alignment Timescales}
The proposed orientation of the grain alignment field places constraints on the previously discussed grain alignment mechanisms. Given the expected toroidally oriented magnetic field of a circumstellar disk, grain alignment through such a field must occur without internal alignment to produce the observed polarization vector orientation in BHB07-11's disk. Additionally, grain alignment through a relative gas--dust velocity flow is dependent on both the direction of the alignment flow and the geometry of the aligned grains. The Badminton Birdie mechanism can produce the observed polarization vector morphology in two ways: effectively prolate grains and a velocity flow along the spiral arms, or effectively oblate grains and a velocity flow perpendicular to the spiral arms. Under these conditions, alignment with either a magnetic field or a relative gas--dust velocity flow can produce the observed polarization. To break this degeneracy we investigate the characteristic alignment timescales for magnetic field and velocity flow induced grain alignment.

Again motivated by the findings of \cite{Looney_L1448_2025}, we first investigated the possibility of grain alignment driven by a relative gas--dust velocity flow in BHB07-11's spiral arms. To determine the prevalence of velocity flow driven grain alignment, we calculated the grain alignment timescale for a large population of oblate and prolate grain configurations. The formalism governing the \textit{v}-MET mechanism is limited to oblate grains, and the quantities needed to calculate the alignment timescale cannot be robustly justified without extensive and complex simulations, meaning that we cannot appropriately constrain the presence of \textit{v}-MET alignment in this work. Badminton Birdie alignment, on the other hand, can be constrained for both prolate and oblate grains with a series of physical parameters that can be readily justified from observations alone. It is important to note that the Badminton Birdie mechanism is built upon the assumptions that gas drag occurs within the Epstein regime, the relative gas--dust velocity flow is subsonic, and that gas reflection off of dust grains is specular rather than diffuse \citep{lin_badminton_2024}. Additionally, for the purposes of the following calculations, we assumed that the effective grain shapes are axisymmetric. For Badminton Birdie alignment, grain geometry is parameterized by:(1) the grain aspect ratio, $\bar{a}\equiv a/c$, where $c$ is the grain length along the axis of symmetry and $a$ is the grain length perpendicular to the axis of symmetry; and (2) the center-of-mass offset ratio, $\bar{g}\equiv g/c$, where $g$ is the length between the geometric center of the ellipsoid and the center of mass. Thus, $\bar{a}\in(0,1)$ and $\bar{a}\in(1,\infty)$ represent prolate grains and oblate grains, respectively. For this analysis we considered $\bar{a}\in[0.1,10]$ and $\bar{g}\in[0,1)$. The characteristic timescale for Badminton Birdie alignment is given by the oscillation damping time, $t_{d,s}$ \citep{lin_badminton_2024}: 
\begin{align}
    t_{d,s}&= \frac{\rho_{s}c}{\rho_{g}v_{\text{th}}}\breve{t}_{d}\label{eq:damping}\\ 
    v_{\text{th}} &= \sqrt{\frac{8kT}{\pi\mu m_{p}}}\label{eq:thermal velocity}\\
    \breve{t}_{d,s}&\equiv\frac{8}{15}\frac{\bar{a}(1+\bar{a}^{2}+5\bar{g}^{2})}{\bar{g}^{2}E[1-x^{2}]+(1-\bar{a}^{2})^{2}E[x^{2}-x^{4}]},\label{eq:damping unitless} 
\end{align}
where $v_{\text{th}}$ is the thermal velocity at temperature $T$, $\breve{t}_{d,s}$ is a dimensionless quantity that is dependent on grain geometry parameters, and $E[1-x^2]$ and $E[x^{2}-x^{4}]$ are integrals which depend on the grain geometry. These integrals are described in appendix section \ref{sec:Analysis Variables}. We use $c=25\,\mu\text{m}$ as the representative grain size and $\rho_{s}=1.675\,\text{g}\,\text{cm}^{-3}$ in accordance with \cite{birnstiel_disk_2018}. We use the Toomre $Q$ parameter to determine the gas mass density, $\rho_{g}$, where $c_{s}$ is the isothermal sound speed, 
\begin{align}
\Omega=\sqrt{GM/R^{3}},
\end{align}
is the Keplerian orbital frequency, $\Sigma$ is the gas surface density, and $H=c_{s}/\Omega$ is the pressure scale height:
\begin{align}
    Q&\equiv\frac{c_{s}\Omega}{\pi G\Sigma}\\
    \rho_{g}&=\frac{\Sigma}{\sqrt{2\pi}H}
\end{align}
Combining these equations results in the gas mass density in terms of known quantities:
\begin{align}
    \rho_{g}=\frac{M}{\pi\sqrt{2\pi}R^{3}Q}.
\end{align}
Using $M=2.25\,\text{M}_{\odot}$ \citep{Alves_nature_2019}, $Q=3.9$ \citep{Alves_nature_2019}, and $R=10\,\text{AU}$, which is the closest distance between BHB07-11B and the spiral arms, we find a gas mass density of $\rho_{g}=4.35\times10^{-11}\,\text{g}\,\text{cm}^{-3}$. We adopt a midplane gas temperature of $T=50\,\text{K}$, which is roughly consistent with the findings of \cite{ober_tracing_2015} and \cite{heese_spread_2017}.

\begin{figure*}
\begin{tabular}{cc}
    \centering
    \includegraphics[width=0.475\textwidth]{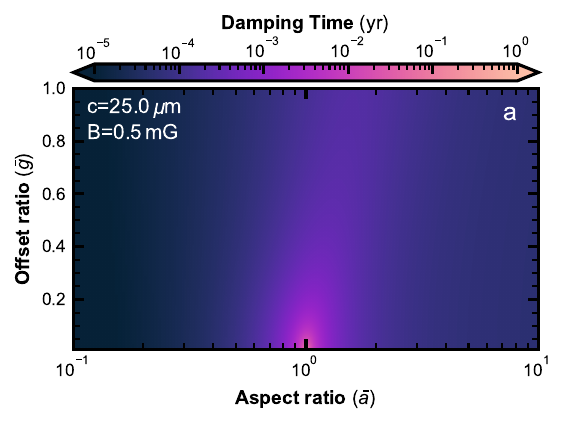} &
    \includegraphics[width=0.475\textwidth]{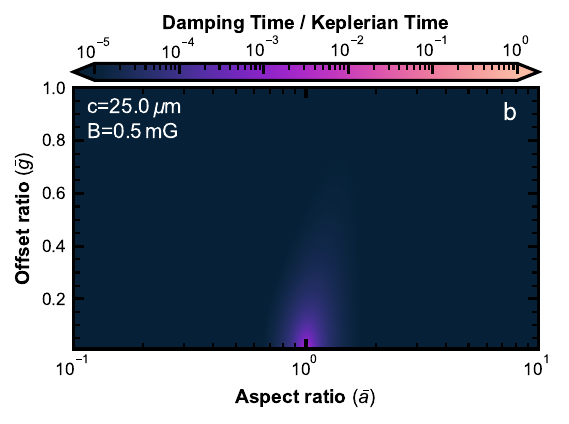} \\
    \includegraphics[width=0.475\textwidth]{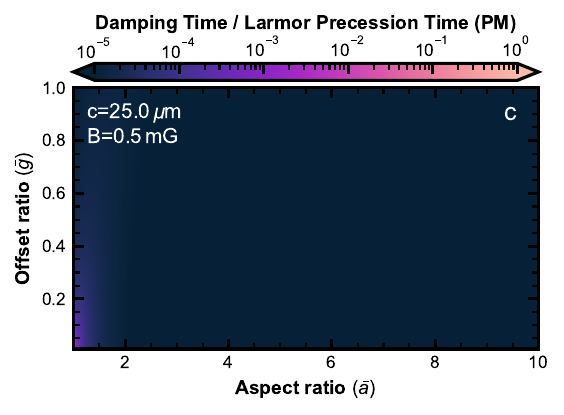} &
    \includegraphics[width=0.475\textwidth]{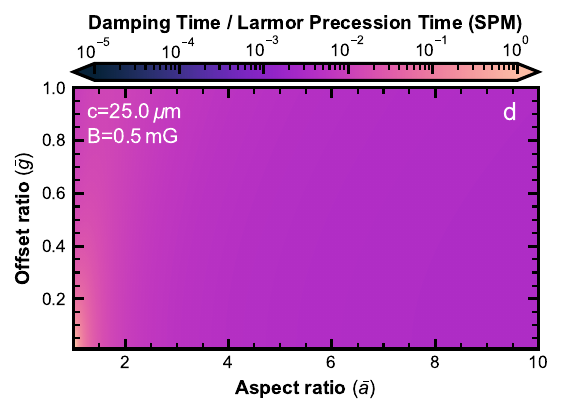}
\end{tabular} 
\caption{Calculations of (a) the grain damping time, (b) the ratio of the damping time to the Keplerian time, and the ratio of the damping time to the Larmor precession time for (c) the paramagnetic and (d) superparamagnetic cases.}
\label{fig:timescale plots}
\end{figure*}

Figure \ref{fig:timescale plots}a shows the damping timescale for the stated prolate and oblate grain configurations using the adopted density, grain size, and temperature values.
To determine the feasibility of Badminton Birdie grain alignment within the spiral arms, we compare the damping time to the Keplerian orbital time:
\begin{align}
    t_{\text{kep}} =\frac{2\pi}{\sqrt{GM/{R^{3}}}}.\label{eq:keplerian time}
\end{align}
If $t_{\text{kep}} < t_{d,s}$ then a given grain is dominated by Keplerian motion and damping is inefficient. If $t_{\text{kep}} > t_{d,s}$, the opposite is the case, as the relative gas--dust velocity flow is able to induce damping torques and enable mechanical grain alignment. To determine the lower limit on the Keplerian time, and thus make the most limiting comparison to the damping time, we assumed that BHB07-11B holds all of the mass within the binary for the purposes of this calculation. Additionally, we performed this calculation using a distance of $R=10\,\text{AU}$, which is approximately the minimum distance between BHB07-11B and the center of the primary spiral arm. These two quantities allow for $t_{\text{kep}}$ to be minimized, and thus place the tightest constraint on mechanical grain alignment. Using these values we find a Keplerian time of $t_{\text{kep}}=21.1\,\text{yr}$. The ratio of the damping time to the Keplerian time for all possible prolate grain configurations is shown in Figure \ref{fig:timescale plots}b. Damping torques are efficient for a majority of prolate and oblate grain configurations, indicating that Badminton Birdie alignment is indeed feasible within BHB07-11's spiral arms. As expected, damping torques are inefficient for nearly spherical grains with small offset ratios.

It is now useful to compare the Badminton Birdie alignment timescale to that of magnetic field alignment (\textit{B}-RAT, \textit{B}-MET), i.e., the Larmor precession time, $t_{\text{L}}$ \citep{tazaki_radiative_2017}:
\begin{align}
    t_{\text{L}}&=\frac{4\pi}{5}\frac{g_{e}\mu_{\text{B}}}{\hbar}\rho_{s}s^{-2/3}a_{\text{eff}}^{2}B^{-1}\chi(0)^{-1}\label{eq:larmor time}\\
    &\approx1.3\,\hat{\rho_{s}}\hat{s}^{-2/3}a_{-5}^{2}\hat{B}^{-1}\hat{\chi}^{-1}\,\text{yr}\label{eq:larmor time simplified}
\end{align}
We adopt a representative magnetic field strength of $0.5\,\text{mG}$ \citep{Khaibrakhmanov_magnetic_2024} and use the previously established values for the grain major and minor axis lengths and material density. Equations \ref{eq:damping} and \ref{eq:larmor time simplified} show that the ratio of the Larmor precession time to the damping time follows $t_{d,s}/t_{L}\propto c^{-1/3}B$, as $t_{L}\propto c^{4/3}B^{-1}$ and $t_{d,s}\propto c$. Thus, the ratio of the two timescales decreases as grain size increases and increase with magnetic field strength. Calculations considering alternate grain size and magnetic field strength values are presented in Appendix \ref{sec:Alternate Parameter Calculations}. The Larmor precession timescale is also dependent on the degree of grain paramagnetic and superparamagnetic inclusions, which are encapsulated in the magnetic susceptibility at zero frequency terms:
\begin{align}
    \chi(0)_{\text{PM}} &= 4.2\times10^{-2}f_{p}\left(\frac{T_{d}}{15\,\text{K}}\right)^{-1} \label{eq:PM chi}
\end{align}
and
\begin{align}
    \chi(0)_{\text{SPM}} &=1.2\times10^{-2}N_{\text{cl}}\phi_{\text{sp}}\left(\frac{T_{d}}{15\,\text{K}}\right)^{-1}, \label{eq:SPM chi}
\end{align}
respectively, In the paramagnetic case, magnetic susceptibility is dependent on the grain paramagnetic atom fraction, $f_{p}$, and temperature, $T_{d}$. For this analysis we adopted a grain paramagnetic atom fraction of $f_{P}=0.1$ \citep{Draine_magnetic_1996, tazaki_radiative_2017}. Superparamagnetic inclusions, specifically in the form of iron nanoclusters, can greatly enhance a grain's magnetic susceptibility, thus altering the Larmor precession timescale \citep{yang_size_2021}. In the superparamagnetic case, magnetic susceptibility is dependent on the number of atoms per nanocluster, $N_{\text{cl}}$, the superparamagnetic atom fraction, $\phi_{\text{sp}}$, and the grain temperature. The number of atoms per nanocluster is expected to be in the range $N_{\text{cl}}=10^{3}-10^{5}$ \citep{Jones_magnetic_1967}, so for this analysis we used $N_{\text{cl}}=10^{4}$ and $f_{p}=0.03$ \citep{Bradley_1994, Martin_1995}. For both cases we adopted a dust temperature of $T_{\text{d}}=29.3\,\text{K}$, in accordance with Figure \ref{fig:Fitting results}b. 

The ratios of the damping time to the Larmor precession time for the paramagnetic and superparamagnetic cases over all prolate grain configurations is shown in Figs. \ref{fig:timescale plots}c and \ref{fig:timescale plots}d, respectively. As with the \textit{v}-MET mechanism, the current formalism governing magnetic field alignment allows only for the consideration of effectively oblate grains; however, the physical quantities necessary for calculating the grain alignment timescale can be robustly justified, allowing for comparison to the Badminton Birdie alignment timescale. In the case of paramagnetic grains, all oblate grain configurations favor mechanical alignment over \textit{B}-RAT alignment, as the damping time is less than the Larmor precession time. In the superparamagnetic case, however, a small fraction of grain configurations $(\bar{a}\lesssim 1.2,\,\bar{g}\lesssim0.1)$ permit magnetic field alignment to dominate. Nevertheless, Badminton Birdie alignment is favorable over magnetic field alignment in both the paramagnetic and superparamagnetic cases, as the damping time is less than the Larmor precession time for a large majority of grain configurations.

\subsubsection{Morphological Constraints}
\label{sec:Morphological Constraints}
The complex formalism of the \textit{k}-RAT mechanism makes alignment timescale calculations challenging without extensive simulations that sample a diverse parameter space. Instead, to constrain grain alignment by an anisotropic radiation field, we developed a fiducial polarization model based on the characteristic circular vector pattern created by \textit{k}-RAT alignment and compare this model to BHB07-11's observed polarization vectors. 

Considering BHB07-11's constituent objects as the two primary radiation sources, we used the $1/d^{2}$ polarization intensity weighting scheme from the \textit{k}-RAT model shown in \cite{Alves_2018}, where $d$ is the distance from a given radiation source. Our model differs from that in \cite{Alves_2018} by incorporating the now resolved positions of BHB07-11A and BHB07-11B as well as the posited \textit{k}-RAT polarization vector pattern from \cite{yang_Does_2019}, both of which were not available at the time of the prior model's creation. Using a total bolometric luminosity of $3.5\,\text{L}_{\odot}$ \citep{Brooke_Spitzer_2007, Alves_nature_2019}, we attributed $1.5\,\text{L}_{\odot}$ to BHB07-11A and $2.0\,\text{L}_{\odot}$ to BHB07-11B based on the measured flux densities of their respective circumstellar disks, which we then used to scale the polarized intensity distribution of our model. To account for this initial assumption in our model, we sampled a luminosity value for BHB07-11A from a Gaussian distribution with $\mu=2\,\text{L}_{\odot}$ and $\sigma=0.1\,\text{L}_{\odot}$, and assigned a corresponding luminosity to BHB07-11B such that the total luminosity of the system matched the reported value. We repeated this process 1000 times to produce the average fiducial polarization model and properly encapsulate the uncertainty in the luminosity selection process. We then determined the absolute angular difference between the model and observed Band 7 polarization vectors, again accounting for the headless nature of polarization vectors. We incorporated the beam sampling method used in the spiral arm alignment model, and again used Gaussian KDE to represent the underlying angular residual distribution.

\begin{figure*}
    \includegraphics[width=\linewidth]{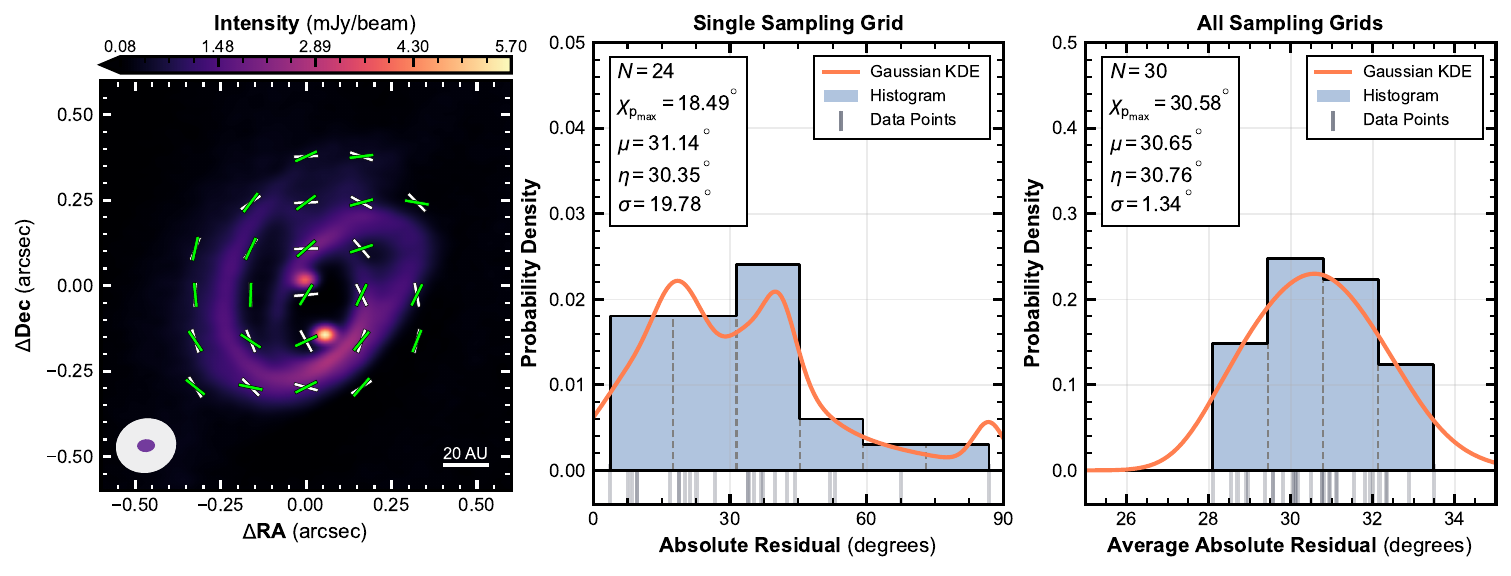}
    \caption{Left: The high-resolution Band 6 dust continuum from in Fig. \ref{fig:Band 6 HR} overlaid with the \textit{k}-RAT model vectors (green) and observed Band 7 polarization vectors (white). Vectors are sampled once per Band 7 beam and are not scaled to the polarization percentage. The displayed vectors represent one of the 30 possible independent sampling grids. The Band 7 (white) and high-resolution Band 6 (purple) beam sizes and position angles on the sky are shown in the bottom left. Center: The distribution of absolute differences between the \textit{k}-RAT model and observed Band 7 polarization vectors shown in the left panel. The bandwidth of the KDE kernel is proportional to the uncertainty in the absolute difference between the model and observed polarization vectors. Right: The distribution of the average absolute residuals from each of the 30 independent sampling grids. Again, the bandwidth of the KDE kernel is proportional to the uncertainty in each average absolute difference measurement. Statistics for the distributions shown in the middle and right panels are in the top left of each panel, respectively.}
    \label{fig:kRAT KDE plot}
\end{figure*}

Figure \ref{fig:kRAT KDE plot} shows a representative residual distribution with the corresponding observed and model polarization vectors, as well as the average value of each of the 30 beam sampled residual distributions. The average absolute residuals for each of the 30 distributions lie between $|\Delta\chi|=28.1^{\circ}$ and $|\Delta\chi|=33.5^{\circ}$, indicating poor agreement between model and observed polarization morphologies. These results are roughly consistent with the \textit{k}-RAT vector morphology model in \cite{Alves_2018}, who found an average residual of $|\Delta\chi|=20.2^{\circ}$ using different model parameters.

Based on the resulting angular residual distributions of the spiral arm and \textit{k}-RAT vector morphology models, we find that the observed polarization vectors are best explained through grain alignment within the spiral arms. Considering the currently available formalism, we find that mechanical grain alignment (Badminton Birdie or \textit{v}-MET) is expected to dominate over alignment with magnetic fields (\textit{B}-RAT, \textit{B}-MET).
\section{Discussion}
\label{sec:Discussion}
\subsection{Morphological Degeneracies}
\begin{figure*}
    \centering
    \includegraphics[width=\linewidth]{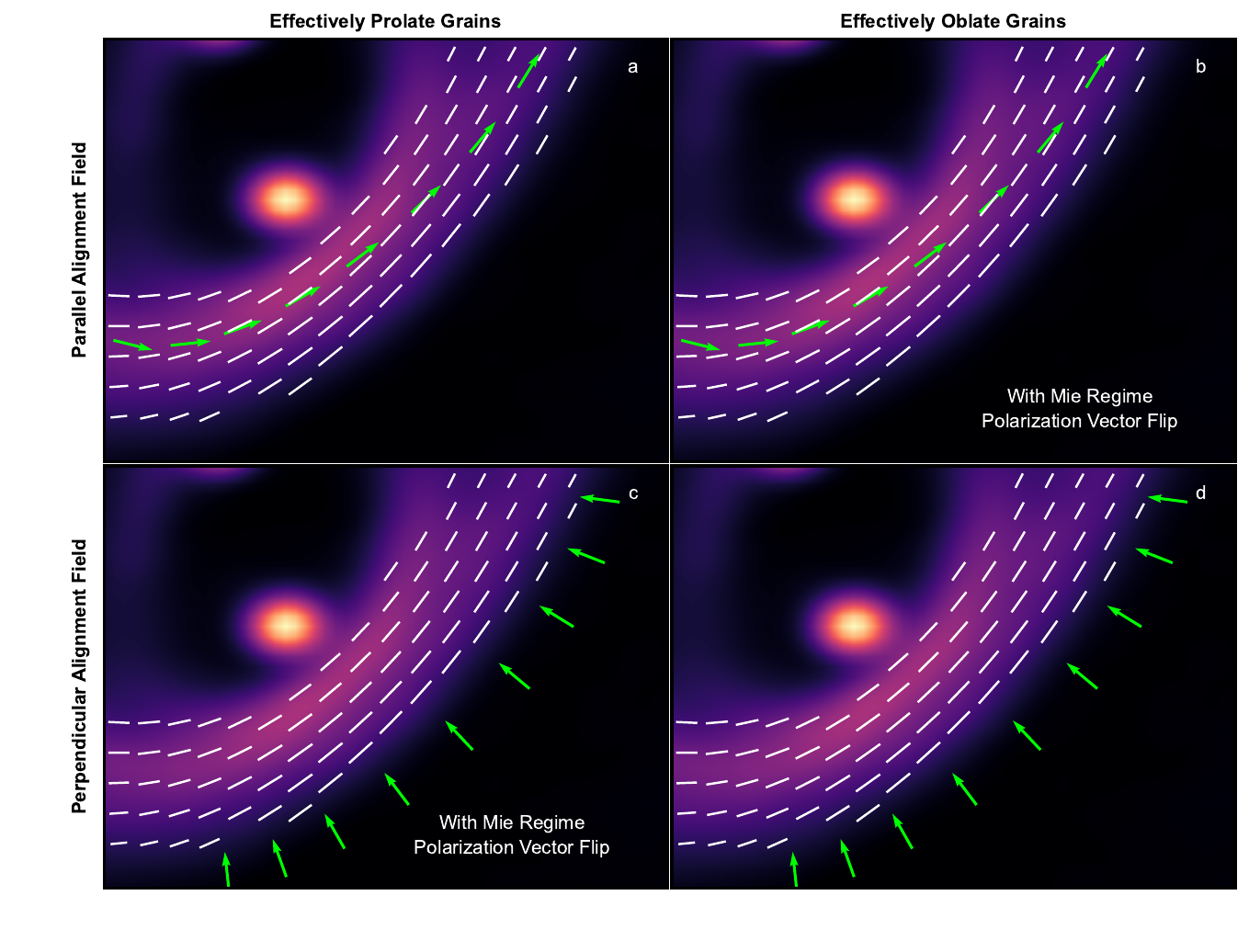}
    \caption{The polarization vector morphologies that can be produced when considering the orientation of the grain alignment field and the effective shape of the emitting grains. The direction of the alignment flow is shown by the green arrows in each panel.}
    \label{fig:Degeneracy diagram field}
\end{figure*}

Although we posit that the observed polarization is due to mechanical grain alignment, there exist degeneracy-inducing mechanisms that limit our understanding of BHB07-11's dust grain population. In this section, we consider possible degeneracies induced by Mie regime vector rotation \citep{guillet_Polarized_2020}, the alignment field orientation, and the effective shape of the emitting dust grains. The combination of these mechanisms creates eight possible cases with two distinct polarization vector morphologies. Four of the eight possible cases result in polarization vectors being rotated by $90^{\circ}$ from their observed orientation, indicating that the combinations of mechanisms that create these vector morphologies are not present in BHB07-11. The remaining four cases produce the observed vector morphology and are discussed below (Fig. \ref{fig:Degeneracy diagram field}).

\subsubsection{Mie Regime Vector Rotation}
Grains in the Mie regime, in which the grain size is comparable to the observation wavelength, can cause polarization vectors to rotate by $90^{\circ}$ from the geometric orientation of the emitting dust grain, thus masking the grain alignment mechanism \citep{guillet_Polarized_2020}. The `flipped' polarization vectors manifest mathematically as negative polarization percentages. \cite{guillet_Polarized_2020} showed that when dust grains are assumed to be composed of astronomical silicates \citep{draine_Optical_1984}, which feature a constant dust opacity index of $\beta=2$, the average polarization percentages for BHB07-11 in Bands 3, 6, and 7 are consistent with large grains ($500 \,\mu\text{m}$) and flipped polarization vectors. 

However, we find that the assumption of a constant dust opacity index does not accurately describe BHB07-11's grains. As discussed in Section \ref{sec:Polarization Spectra} we found an average beam sampled dust opacity index of $\beta=1.53\pm0.07$, which is notably shallower than the value used in \cite{guillet_Polarized_2020}. The dust opacity index also shows clear spatial variation throughout the disk (Fig. \ref{fig:Beta map}). Additionally, the average spectral index is close to that of the ISM 
\citep[$\beta_{\text{ISM}}=1.7$,][]{Weingartner_and_Draine_ISM}, indicating that the grains in BHB07-11's disk are far smaller than the $500\,\mu\text{m}$ required to induce the Mie regime polarization flip \citep{Li_ISM_distribution}. Although degeneracy in the dust opacity index exists due to its dependence on grain composition and shape, this parameter is considerably more sensitive to grain size \citep{Draine_sensitivity_2006}, and is therefore a reliable metric for differentiating between large and small grains. 

\subsubsection{Alignment Field Orientation}
Additional morphological degeneracy stems from the orientation of the grain alignment field with respect to the spiral arms and the from effective geometry of the aligned grain population. If the alignment field direction is parallel to the spiral arms (i.e., follows the curvature of the arms), then the observed polarization vectors can only be produced by effectively prolate dust grains. However, if the relative gas--dust velocity flow is instead perpendicular to the spiral arms (i.e., depositing material onto the spiral arms), the observed polarization vectors can only be produced with effectively oblate grains. To determine the orientation of the alignment field, it is necessary to understand the velocity components of the gas and dust near the disk midplane. Although dense gas tracers can provide useful insights into line-of-sight gas kinematics near the disk midplane, it is not possible to directly recover similar information on dust kinematics through observations. Although simulations of spiral arm formation in circumstellar disks suggest that dust can flow along the spiral arms as opposed to perpendicularly \citep{rowther_role_2024}, we cannot determine which combination of velocity flow orientation and effective grain shape produce the observed polarization from an observational standpoint. Nevertheless, it is still clear that grain alignment within the spiral arms is likely the result of the Badminton Birdie mechanism. Although the orientation of the relative gas--dust velocity flow throughout the disk may actually be more complex than simply parallel or perpendicular to the spiral arms, we have illustrated that the Badminton Birdie mechanism can explain the observed polarization vectors given such a field orientation.

\subsection{Coupling of Gas and Dust}
\begin{figure}
    \centering
    \includegraphics[width=\linewidth]{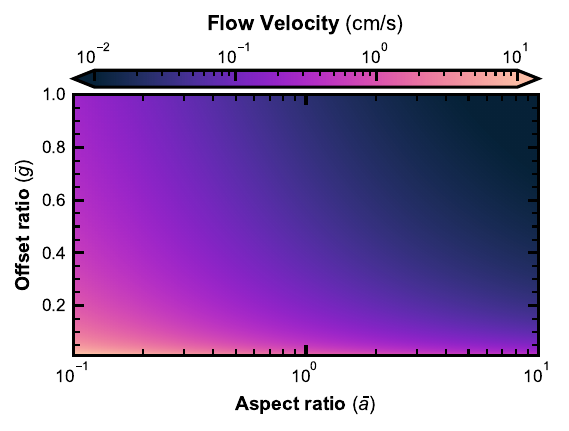}
    \caption{The gas--dust flow velocity required to induce Badminton Birdie alignment for all prolate grain configurations with thermal angular velocities.}
    \label{fig:Flow velocity}
\end{figure}

The Badminton Birdie mechanism necessitates the existence of a relative velocity flow between the gas and emitting dust grains in the disk. To determine the necessary drift velocity to induce Badminton Birdie alignment, we use the following equations from \cite{lin_badminton_2024}:
\begin{align}
    A_{t,s}&\equiv \frac{fv_{\text{th}}}{nc^{3}}\breve{A}_{t,s}\\
    \breve{A}_{t,s}&\equiv\frac{1}{16\bar{a}\bar{g}E[1-x^{2}]},
\end{align}
where $A_{t,s}$ is the threshold drift velocity, $f$ is a multiplicative factor parameterizing grain rotational energy given by external factors, such as RATs, that can kick grains out of stable alignment, and $\breve{A}_{t,s}$ is a dimensionless factor encompassing grain geometry. We adopted $f=3/2$ as a representative value for grains in thermal equilibrium due to RATs \citep[e.g.,][]{tazaki_radiative_2017}. Figure \ref{fig:Flow velocity} shows the required flow velocities for Badminton Birdie alignment for a range of prolate and oblate grain configurations using the representative grain size of $c=25\,\mu\text{m}$. A majority of grain configurations require drift velocities on the order of $\sim1\,$cm s$^{-1}$, with the maximum required drift velocity for the sampled grain configurations being $\sim22$ cm s$^{-1}$. These velocities are generally representative of gas--dust drift velocities in circumstellar disks \citep{birnstiel_dust_2024}. 

In addition to the required drift velocity, gas--dust coupling plays an important role in determining the presence of a relative gas--dust velocity flow. If gas and dust are highly coupled, a relative gas--dust velocity flow is far more difficult to achieve in comparison to the case of poor coupling. The degree to which dust grains are coupled to gas is given by the Stokes number, St, which is primarily dependent on grain size \citep{batygin_self-consistent_2022}. The inferred $\sim10\text{ - }50\,\mu\text{m}$ sized grains should correspond to a small Stokes number $(\text{St}\ll1)$, indicating a high degree of gas coupling and a diminished relative gas--dust velocity flow. We suspect that RATs may not be strong enough to drive grains to thermal angular velocities and would allow even the small grains of diminished relative velocity to become aligned. This scenario is at least consistent with our result that the RAT alignment is not responsible for the observed polarization morphology. Additionally, given the low required drift velocities for all grain configurations, Badminton Birdie alignment may still occur even with the inferred small grain size. 
\subsection{Caveats}
Here we discuss the observational limitations that affect this work. The most impactful limitation is the lack of high-resolution full polarization mode observations of BHB07-11 (i.e., a beam size able to resolve the spiral arms). Such observations are critical in imaging complex polarization morphologies and in identifying grain alignment mechanisms \citep[e.g.,][]{Stephens_Aligned_2023, Looney_L1448_2025}. High-resolution observations would directly reveal the degree of vector alignment within the spiral arms, as well as how polarization changes within the individual circumstellar disks of BHB07-11A and BHB07-11B. In particular, such observations might reveal previously unseen scattering morphology on such scales, thus providing valuable information on grain population sizes and distributions. The three-dimensional orientation of the spiral arm structure is also not known due to the nature of astronomical observations, leading to uncertainty in the proposed relative gas--dust velocity flow direction and resulting polarization model. Nevertheless, we believe this work provides the most complete and comprehensive analysis of BHB07-11's polarization mechanism to date.

There also exist limitations in the theoretical and computational framework that motivates our comparison of grain alignment timescales. First, it is possible that grains are aligned via a magnetic field, and that the aligning magnetic field is too complicated to model efficiently and effectively. Despite the grain geometry constraints shown in Section \ref{sec:Alignment Timescales}, which effectively constrain magnetic field alignment to a small subset of grain configurations with appropriate (super)paramagnetic inclusions, such alignment is still possible with a particular set of grains. 

Considering \textit{k}-RAT grain alignment, the RAT precession timescale is given approximately by \citep{tazaki_radiative_2017}
\begin{align}
    t_{\text{rad,p}}&\approx1.1\times10^{2}\hat{\rho_{s}}^{1/2}\hat{s}^{-1/3}a_{-5}^{1/2}\hat{T_{d}}^{1/2} \nonumber\\
    &\times\left(\frac{u_{\text{rad}}}{u_{\text{ISRF}}}\right)^{-1}\left(\frac{\bar{\lambda}}{1.2\,\mu\text{m}}\right)^{-1}\left(\frac{\gamma|\overline{\bm{Q_{\Gamma}}}|}{0.01}\right)^{-1}\,\text{yr}, \label{eq:k-RAT time}
\end{align}
where $u_{\text{rad}}$ is the radiation energy density, $\gamma$ is the anisotropy parameter, and $\bar{\lambda}$ and $\overline{\bm{Q_{\Gamma}}}$ are the spectrum-averaged radiation wavelength and RAT efficiency, respectively. Although the normalized grain properties are straightforward to calculate, it is difficult to estimate and robustly justify the radiation field parameters without extensive and complex modeling. Such work falls outside the scope of this work, and thus we can only constrain \textit{k}-RAT grain alignment through the morphological evidence presented in Section \ref{sec:Morphological Constraints}. With respect to the \textit{k}-RAT morphology model, the bulk of photons imparting RATs onto the emitting dust grains within the spiral arms may have been reprocessed from the optical and near-IR into the mid- or far-IR, meaning that our assumption of BHB07-11A and BHB07-11B being the primary radiation sources is not appropriate. Although this scenario is possible, generating a model polarization morphology would require in-depth radiative transfer simulations, which falls well outside the scope of this work.

It is similarly difficult to constrain the \textit{v}-MET alignment timescale, which is given by \citep{hoang_Internal_2022}
\begin{align}
    t_{\text{MET}}&\simeq1.7\times10^{-2}\frac{\Omega/\Omega_{T}}{a_{-5}^{\,1/2}Q_{\text{prec}}s_{d,-1}^{2}}\left(\frac{1}{n_{8}T_{1}^{1/2}}\right)\,\text{yr}\label{eq:v-MET time}\\
    Q_{\text{prec}}&=se(e^{2}-1)K(\Theta,e)\sin(2\Theta),\label{eq:v-MET efficiency}
\end{align}
where $\Omega/\Omega_{T}$ is the ratio of the grain angular momentum to the thermal angular momentum, $Q_\text{prec}$ is the precession efficiency, and $s_{d}$ is the drift velocity ratio. The formalism for the \textit{v}-MET alignment mechanism is based on oblate grains, meaning that equations \ref{eq:v-MET time} and \ref{eq:v-MET efficiency} must be adapted to prolate grains in order to be properly used in this analysis. Future work may enable such analysis, but at this time it is not possible to robustly constrain the \textit{v}-MET alignment timescale in the framework of prolate grains.

\subsection{Future Developments}
This work serves as motivation for future developments in grain alignment theory. In particular, the formalism for RAT/MET alignment is limited to oblate grains, whereas observations have shown that effectively prolate grains are required to explain the observed polarization in HL Tau \citep{yang_Does_2019, mori_modeling_2021, Stephens_Aligned_2023, lin_Panchromatic_2024} and other sources \citep[e.g.,][]{sadavoy_dust_2019, harrison_dust_2019}. Such developments would enable grain alignment mechanisms to be more robustly constrained in circumstellar disks. Additionally, simulation-focused work on the presence of relative gas--dust velocity flows in circumstellar disks would further elucidate grain alignment mechanisms, more specifically the prevalence of Badminton Birdie alignment.

\section{Conclusions}
\label{sec:Conclusions}
We have presented an analysis of the Class I protobinary system BHB07-11 using ALMA Band 3, 6, and 7 polarization mode dust continuum observations in combination with high-resolution Band 6 dust continuum and molecular line observations. Multi-wavelength observations allow for BHB07-11's polarization mechanism to be constrained, which sheds new light on the star formation and dust grain growth processes. Our primary conclusions are:
\begin{enumerate}
    \item The Band 3, 6, and 7 polarization mode dust continuum observations reveal strikingly similar polarization vector morphologies at all observed wavelengths. Polarization vectors curl azimuthally around the center of the system, and the polarized intensity maps closely to the shape of the Stokes \textit{I} emission in each band. The high-resolution Band 6 dust continuum observations highlight BHB07-11's spiral arm structures, which overlap one another to form the `cosmic pretzel'. When overlaid on the high-resolution Band 6 images, polarization vectors are well aligned with the spiral arms, suggesting that the polarization mechanism is directly related to the complex dust morphology.
    \item Polarization vectors are consistent with emission from aligned dust grains rather than from dust self-scattering. We compared magnetic field alignment, radiation-induced alignment, and velocity flow alignment to determine the dominant alignment mechanism. Through a series of morphological models and alignment timescale calculations, we find that the Badminton Birdie mechanism is most likely responsible for the observed polarization.
    \item Using the methods described in \cite{Hildebrand_Davidson_Dotson_Dowell_Novak_Vaillancourt_2000}, we made estimates of intrinsic polarization, dust temperature, and optical depth throughout the disk. The optical depth estimates enabled the spatial distribution of the dust opacity spectral index to be determined. We found a beam sampled average dust opacity spectral index of $\langle\beta\rangle=1.53\pm0.07$, indicating that grains have not undergone significant growth from ISM-scale grains. We estimate that grains have sizes on the order of $\sim10\text{--}50\,\mu\text{m}$ based on the dust opacity spectral index and the observed polarization.
    \item We carefully investigated possible degeneracy-inducing mechanisms in relation to the observed polarization morphology. In contrast to \cite{guillet_Polarized_2020}, we demonstrated that the polarization vectors are not subject to the Mie regime vector rotation, and trace the orientation of the emitting grains. Given this result, we conclude that the observed polarization can be explained by two cases: effectively prolate grains and a velocity flow along the spiral arms, and effectively oblate grains and a velocity flow onto the spiral arms. Additional observations and simulations are required to determine which scenario is correct.
\end{enumerate}

\section*{Acknowledgments}
This paper makes use of the following ALMA data: ADS/JAO.ALMA$\#$2013.1.00291.S and ADS/JAO.ALMA$\#$2016.1.01186.S. ALMA is a partnership of ESO (representing its member states), NSF (USA) and NINS (Japan), together with NRC (Canada), NSTC and ASIAA (Taiwan), and KASI (Republic of Korea), in cooperation with the Republic of Chile. The Joint ALMA Observatory is operated by ESO, AUI/NRAO and NAOJ. The National Radio Astronomy Observatory and Green Bank Observatory are facilities of the U.S. National Science Foundation operated under cooperative agreement by Associated Universities, Inc.

L.W.L. acknowledges support from NSF AST-1910364 and NSF AST-2307844.
Z.-Y.D.L. acknowledges support from NASA 80NSSC18K1095, the Jefferson Scholars Foundation, the NRAO ALMA Student Observing Support (SOS) SOSPA8-003, the Achievements Rewards for College Scientists (ARCS) Foundation Washington Chapter, the Virginia Space Grant Consortium (VSGC), and UVA research computing (RIVANNA).
ZYL is supported in part by NASA 80NSSC20K0533
and NSF AST-2307199.

A.F. acknowledges Dr. Natalie Butterfield and the NRAO data support staff for their assistance with data reduction, as well as Shurui Lin and Padma Venkatraman for their advice regarding statistical analysis and visualization.

\vspace{5mm}
\facility{ALMA}
\software{astropy \citep{astropy}, CARTA \citep{CARTA}, CASA \citep{TheCasaTeam_2022}, cmasher \citep{cmasher}, KDEpy \citep{KDEpy}, matplotlib \citep{matplotlib}, numpy \citep{numpy}, radfil \citep{2018ApJ...864..152Z}, scipy \citep{scipy}}

\appendix
\restartappendixnumbering
\section{Analysis Variables and Equations}
\label{sec:Analysis Variables}

\begin{table*}
\centering
    \caption{Alignment timescale variables (interpreted from \cite{hoang_Internal_2022})}
    \label{tab:Timescale variables}
    \begin{tabular}{ccccc}
    \toprule
    Use&Symbol&Description&Value/Definition&Equation used\\
    \hline
    \multirow{6}{*}{Physical constants}&$m_{p}$&Proton mass&&Eq. \ref{eq:thermal velocity}\\
    &$k$&Boltzmann constant&&Eq. \ref{eq:thermal velocity}\\
    &G&Gravitational constant&&Eq. \ref{eq:keplerian time} \\
    &$\mu_{\text{B}}$&Bohr magneton&&Eq. \ref{eq:larmor time} \\
    &$g_{\text{e}}$&g-factor&&Eq. \ref{eq:larmor time} \\
    &$\hbar$&Reduced Planck constant&&Eq. \ref{eq:larmor time} \\
    \hline
    \multirow{7}{*}{Environmental variables}&$M$&Stellar mass&$2.25\,\text{M}_{\odot}$&Eq. \ref{eq:keplerian time}\\
    &$R$&Gravitational distance&$10\,\text{AU}$&Eq. \ref{eq:keplerian time}\\
    &$B$&Magnetic field strength$^{1}$&$0.5\,\text{mG}$&Eq. \ref{eq:larmor time}\\
    &$T$&Gas temperature$^{2}$&$50\,\text{K}$&Eq. \ref{eq:thermal velocity}\\
    &$T_{d}$&Dust temperature&$29.3\,\text{K}$&Eqs. \ref{eq:PM chi}, \ref{eq:SPM chi}\\
    &$\rho_{g}$&Gas mass density&$3.8\times10^{-13}\,\text{g}\,\text{cm}^{-3}$&Eq. \ref{eq:keplerian time}\\
    &$n$&Gas number density$^{3}$&$10^{11}\,\text{cm}^{-3}$& \\
    \hline
    \multirow{10}{*}{Grain parameters\tablenotemark{a}}&$a_{1},a$&Grain minor axis&$5\,\mu\text{m}$&$s, \bar{a}$ \\
    &$a_{2},c$&Grain major axis&$25\,\mu\text{m}$&$s,\bar{a},\bar{g}$, Eq. \ref{eq:damping}\\
    &$a_{\text{eff}}$&Grain effective radius&$a_{\text{eff}}^{3}\equiv a_{1}a_{2}^{2}$&Eq. \ref{eq:larmor time}\\
    &$s,\bar{a}$&Aspect ratio&$s=a_{1}/a_{2},\,\bar{a}\equiv a/c$&Eqs. \ref{eq:damping unitless}\\
    &$g$&Offset length&&$\bar{g}$\\
    &$\bar{g}$&Offset ratio&$\bar{g}\equiv g/c$&Eq. \ref{eq:damping unitless}\\
    &$\rho_{s}$&Grain material density$^{4}$&$1.675\,\text{g}\,\text{cm}^{-3}$&Eqs. \ref{eq:damping},\ref{eq:larmor time}\\
    &$f_{p}$&Paramagnetic atom fraction$^{5}$&$0.1$&Eq. \ref{eq:PM chi}\\
    &$N_{\text{cl}}$&Atoms per superparamagnetic cluster$^{6}$&$10^{4}$&Eq. \ref{eq:SPM chi}\\
    &$\phi_{\text{sp}}$&Superparamagnetic atom fraction$^{7}$&$0.03$&Eq. \ref{eq:SPM chi}\\
    \hline    
    \multirow{5}{*}{\textit{k}-RAT parameters}&$u_\text{rad}$&Radiation field energy spectrum&&Eq. \ref{eq:k-RAT time}\\
    &$u_{\text{ISRF}}$&Interstellar radiation field energy spectrum&&Eq. \ref{eq:k-RAT time} \\
    &$\bar{\lambda}$&Spectrum averaged radiation wavelength&See \cite{tazaki_radiative_2017}&Eq. \ref{eq:k-RAT time} \\
    &$\overline{\bm{Q_{\Gamma}}}$&Spectrum averaged RAT efficiency&See \cite{tazaki_radiative_2017}&Eq. \ref{eq:k-RAT time} \\
    &$\gamma$&Anisotropy parameter&See \cite{tazaki_radiative_2017}&Eq. \ref{eq:k-RAT time} \\
    \hline
    \multirow{8}{*}{\textit{v}-MET parameters\tablenotemark{b}}&$I_{||}$&Principal moment of inertial (parallel)&$I_{||}=(8\pi/15)\rho_{s}sa^{5}$&$\Omega_{T}$ \\
    &$\Omega$&Grain angular momentum&&Eq. \ref{eq:v-MET time}\\
    &$\Omega_{T}$&Thermal angular momentum&$\Omega_{T}=\sqrt{kT/I_{||}}$&Eq. \ref{eq:v-MET time}\\
    &$e$&Ellipticity&$e=\sqrt{1-(c^{2}/a^{2})}$&Eq. \ref{eq:v-MET efficiency}\\
    &$\Theta$&Minor axis -- flow alignment angle&&Eq. \ref{eq:v-MET efficiency}\\
    &$K(\Theta,e)$&Fitting function (order unity)&&Eq. \ref{eq:v-MET efficiency}\\
    &$Q_{\text{prec}}$&MET precession efficiency&&Eq. \ref{eq:v-MET time}\\
    &$v_{d}$&Drift velocity&&$s_{d}$\\
    &$s_{d}$&Drift velocity ratio&$s_{d}=v_{d}/v_{\text{th}}$&\\
    \hline
    \multirow{9}{*}{Normalization factors}&$\hat{\rho_{s}}$&Grain material density normalization&$\hat{\rho_{s}} = \rho_{s}/3\,\text{g}\,\text{cm}^{-3}$&Eqs. \ref{eq:larmor time simplified}, \ref{eq:k-RAT time} \\
    &$\hat{s}$&Aspect ratio normalization&$\hat{s}=s/0.5$&Eqs. \ref{eq:larmor time simplified}, \ref{eq:k-RAT time} \\
    &$a_{-5}$&Effective radius normalization&$a_{-5}=a_{\text{eff}}/10^{-5}\,\text{cm}$ &Eqs. \ref{eq:larmor time simplified}, \ref{eq:k-RAT time}\\
    &$\hat{B}$&Magnetic field strength normalization&$\hat{B}=B/5\,\mu\text{G}$&Eq. \ref{eq:larmor time simplified} \\
    &$\hat{\chi}$&Magnetic susceptibility normalization&$\hat{\chi}=\chi(0)/10^{-4}$&Eq. \ref{eq:larmor time simplified} \\
    &$\hat{T_{d}}$&Dust temperature normalization&$\hat{T_{d}}=T_{d}/15\,\text{K}$&Eq. \ref{eq:k-RAT time} \\
    &$s_{d,-1}$&Drift velocity ratio normalization&$s_{d,-1}=s_{d}/0.1$&Eq. \ref{eq:v-MET time}\\
    &$n_{8}$&Gas number density normalization&$n_{8}=n/10^{8}$&Eq. \ref{eq:v-MET time}\\
    &$T_{1}$&Gas temperature normalization&$T_{1}=T/10\,\text{K}$&Eq. \ref{eq:v-MET time}\\
    \bottomrule
    \end{tabular}
    \tablenotetext{a}{Multiple symbols are used to maintain consistency with source material.}
    \tablenotemark{b}{The theoretical basis for \textit{v}-MET alignment is in the framework of oblate grains, so for this section $a$ is the grain major axis and $c$ is the minor axis in accordance with \cite{hoang_Internal_2022}.}
    \tablenotetext{}{Value citations are as follows: 1) \cite{Khaibrakhmanov_magnetic_2024} 2) \cite{ober_tracing_2015, heese_spread_2017} 3) \cite{tomida_grand-design_2017} 4) \cite{birnstiel_disk_2018} 5) \cite{Draine_magnetic_1996} 6) \cite{Jones_magnetic_1967}} 7) \cite{Bradley_1994, Martin_1995}
\end{table*}

The variables and values used in this analysis are listed in Table \ref{tab:Timescale variables}. Additionally, the integrals used in Eq. \ref{eq:damping unitless} are given below; given the linearity of integrals, $E[1-x^{2}] = E[1]-E[x^{2}]$ and $E[x^{2}-x^{4}] = E[x^{2}]-E[x^{4}]$.

\begin{align}
e&\equiv \sqrt{1-\bar{a}^{2}}, \text{for }\bar{a}<1\\
\varepsilon&\equiv \sqrt{\bar{a}^{2}-1}, \text{for }\bar{a}>1\\
E[1]&=
    \begin{cases}
    \frac{2 \arcsin(e)}{e},\,&\text{when } \bar{a} < 1 \\
    2,\,&\text{when }\bar{a} = 1 \\
\frac{2 \operatorname{arcsinh}(\varepsilon)}{\varepsilon},&\text{when }\bar{a} > 1
    \end{cases}\\
E[x^{2}]&=
    \begin{cases}
    -\frac{\sqrt{1-e^{2}}}{e^{2}}+\frac{\arcsin(e)}{e^{3}}, &\text{when } \bar{a} < 1 \\
    \frac{2}{3},\,&\text{when }\bar{a} = 1 \\
    \frac{\sqrt{1+\varepsilon^{2}}}{\varepsilon^{2}}-\frac{\operatorname{arcsinh}(\varepsilon)}{\varepsilon^{3}},&\text{when }\bar{a} > 1
    \end{cases}\\
E[x^{4}]&=
    \begin{cases}
    -\frac{(2e^{2}+3)\sqrt{1-e^{2}}}{4e^{4}}+\frac{3\arcsin(e)}{4e^{5}}, &\text{when } \bar{a} < 1 \\
    \frac{2}{5},\,&\text{when }\bar{a} = 1 \\
    \frac{(2\varepsilon^{2}-3)\sqrt{1+\varepsilon^{2}}}{4\varepsilon^{4}}-\frac{3\operatorname{arcsinh}(\varepsilon)}{4\varepsilon^{5}},&\text{when }\bar{a} > 1
    \end{cases}
\end{align}

\section{Alternative Parameter Calculations}
\label{sec:Alternate Parameter Calculations}
To account for the uncertainty in the grain size and magnetic field values chosen in Section \ref{sec:Alignment Timescales}, we present grain alignment timescale calculations considering larger grain sizes $(100\,\mu\text{m},\,1000\,\mu\text{m})$ and magnetic field strength $(5\,\text{mG})$. For the magnetic field strength used in the main text $(0.5\,\text{mG})$ Badminton Birdie alignment is still favored for a large majority of grain configurations at $c=100\,\mu\text{m}$ and $c=1000\,\mu\text{m}$. On the other hand, the higher magnetic field strength enables magnetic field alignment, particularly with superparamagnetic inclusions, for a larger number of grain configurations in comparison to the value used in the main text. For $c=25\,\mu\text{m}$ grains with superparamagnetic inclusions, magnetic field alignment can occur for $\bar{a}\lesssim 1.5,\,\bar{g}\lesssim0.2$, compared to $\bar{a}\lesssim 1.2,\,\bar{g}\lesssim0.1$ for $B=0.5\,\text{mG}$. Nevertheless, Badminton Birdie alignment is still favorable for superparamagnetic grains in a $B=5\,\text{mG}$ magnetic field.
\begin{figure*}
\begin{tabular}{cc}
    \centering
    \includegraphics[width=0.475\textwidth]{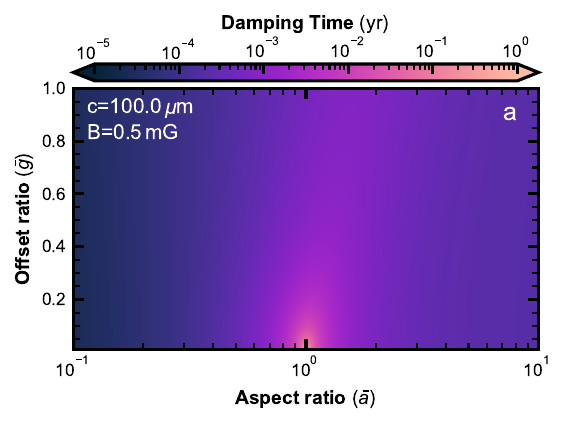} &
    \includegraphics[width=0.475\textwidth]{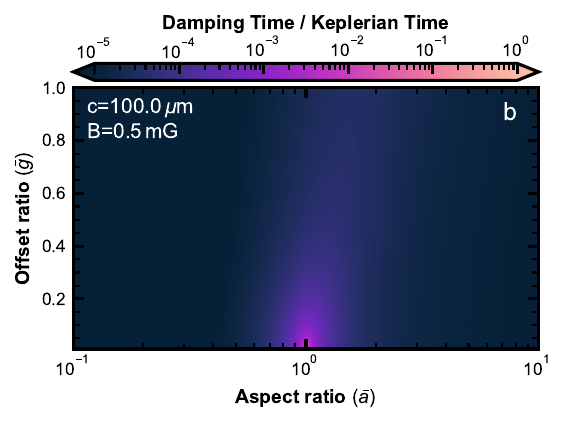} \\
    \includegraphics[width=0.475\textwidth]{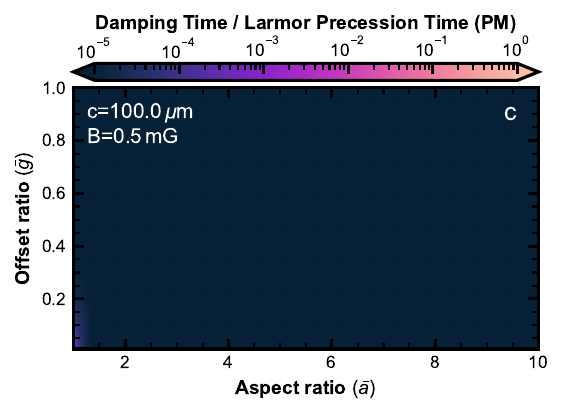} &
    \includegraphics[width=0.475\textwidth]{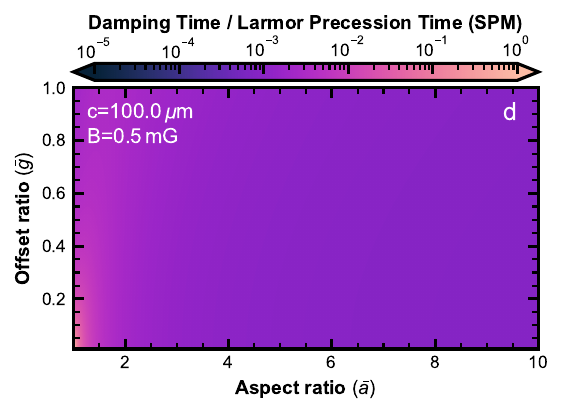}
\end{tabular}
    \caption{The timescale calculations shown in Figure \ref{fig:timescale plots} but using $c=100\,\mu\text{m}$.}
\end{figure*}

\begin{figure*}
\begin{tabular}{cc}
    \centering
    \includegraphics[width=0.475\textwidth]{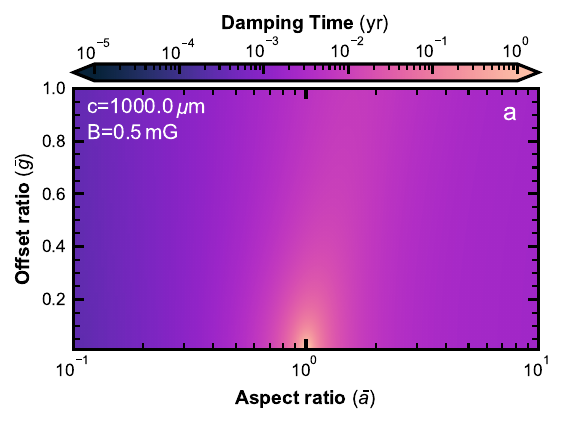} &
    \includegraphics[width=0.475\textwidth]{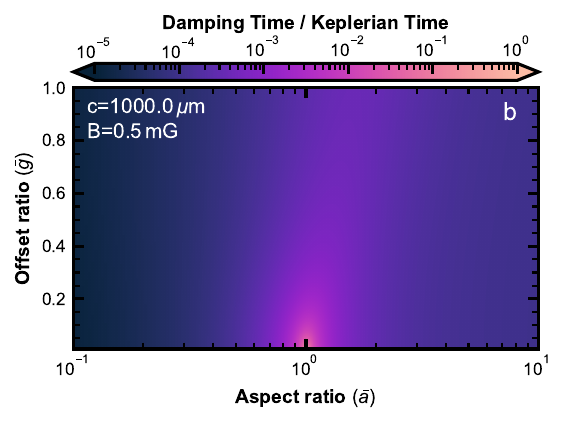} \\
    \includegraphics[width=0.475\textwidth]{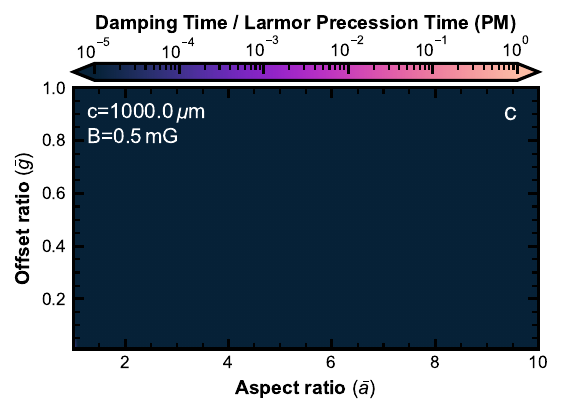} &
    \includegraphics[width=0.475\textwidth]{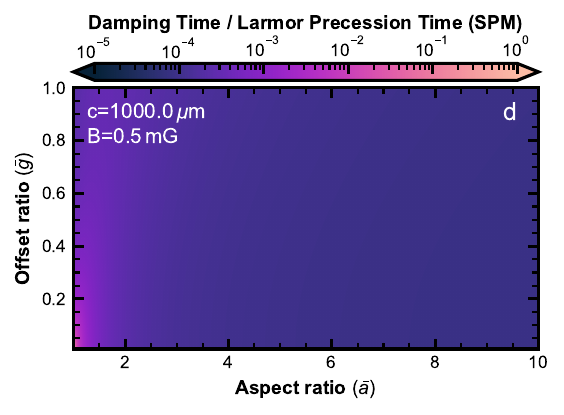}
\end{tabular}
    \caption{The timescale calculations shown in Figure \ref{fig:timescale plots} but using $c=1000\,\mu\text{m}$.}
\end{figure*}

\begin{figure*}
\begin{tabular}{cc}
    \centering
    \includegraphics[width=0.475\textwidth]{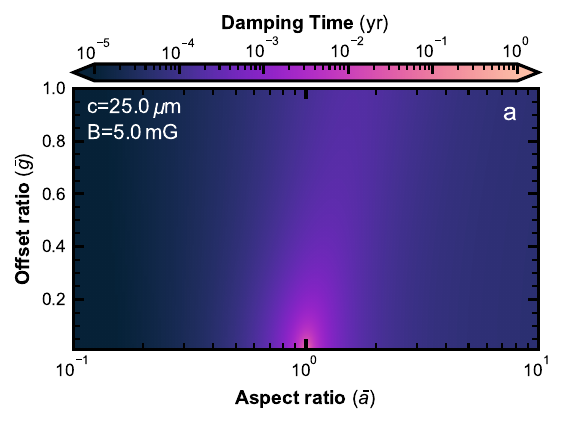} &
    \includegraphics[width=0.475\textwidth]{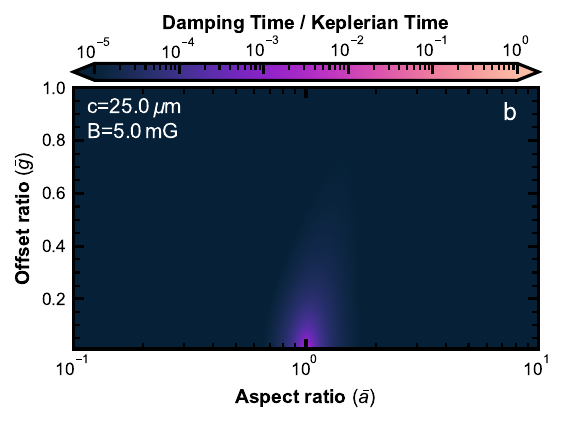} \\
    \includegraphics[width=0.475\textwidth]{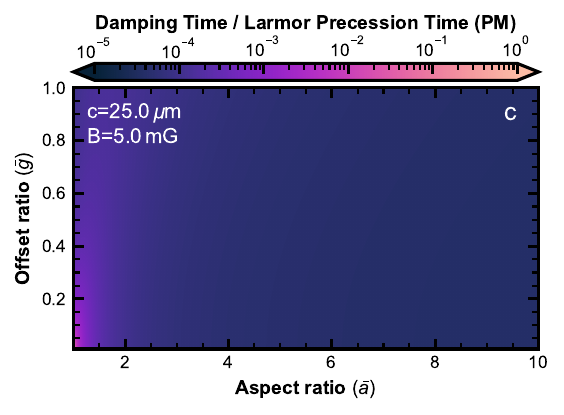} &
    \includegraphics[width=0.475\textwidth]{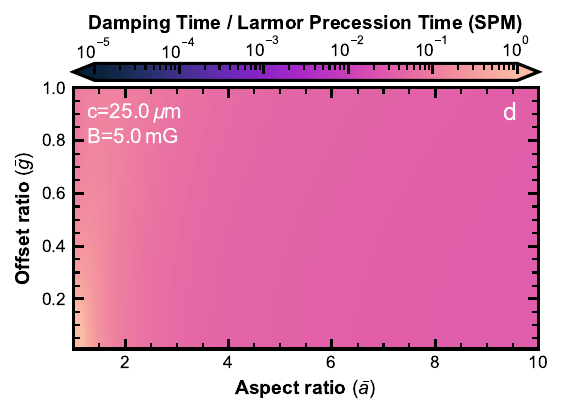}
\end{tabular}
    \caption{The timescale calculations shown in Figure \ref{fig:timescale plots} but using $B=5\,\text{mG}$.}
\end{figure*}

\begin{figure*}
\begin{tabular}{cc}
    \centering
    \includegraphics[width=0.475\textwidth]{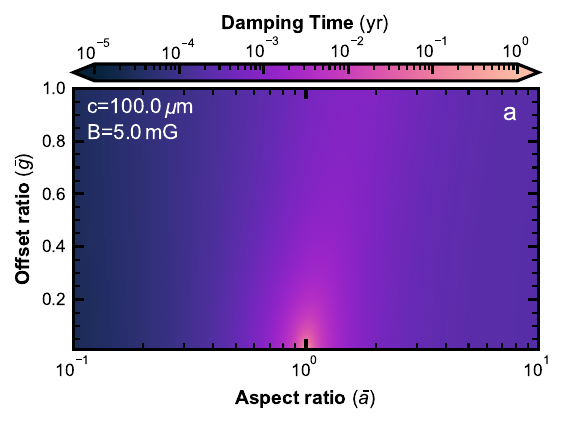} &
    \includegraphics[width=0.475\textwidth]{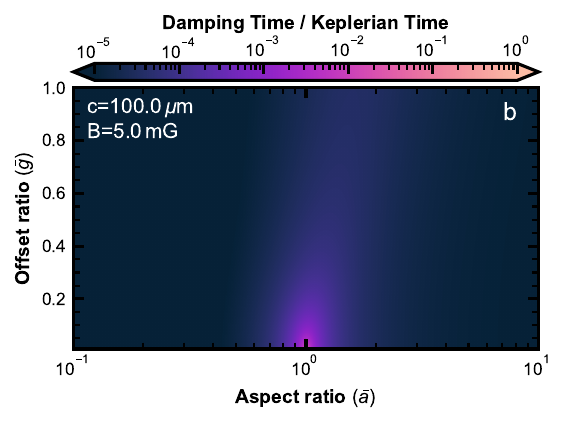} \\
    \includegraphics[width=0.475\textwidth]{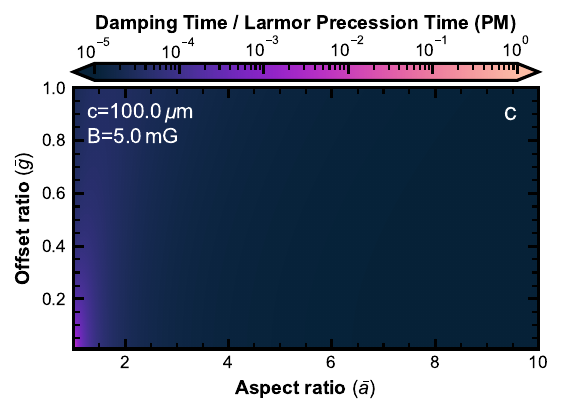} &
    \includegraphics[width=0.475\textwidth]{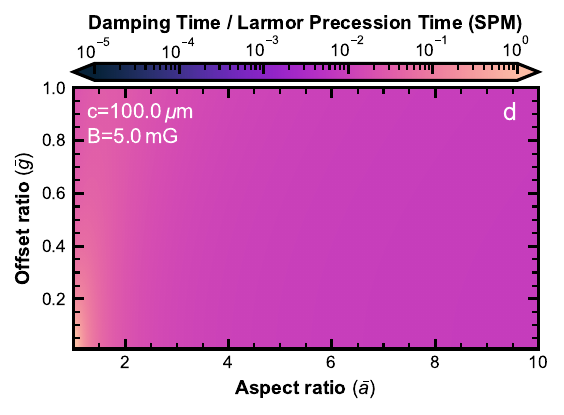}
\end{tabular}
    \caption{The timescale calculations shown in Figure \ref{fig:timescale plots} but using $c=100\,\mu\text{m}$ and $B=5\,\text{mG}$.}
\end{figure*}

\begin{figure*}
\begin{tabular}{cc}
    \centering
    \includegraphics[width=0.475\textwidth]{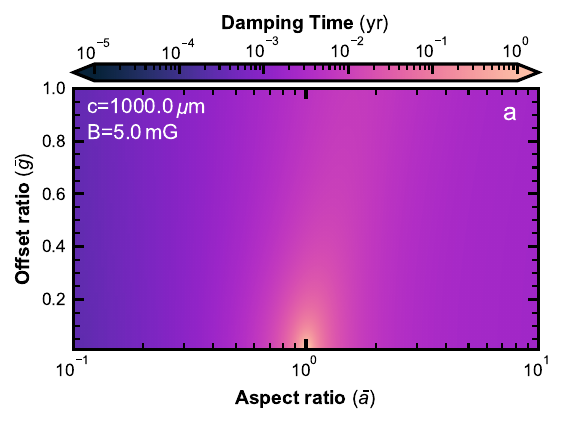} &
    \includegraphics[width=0.475\textwidth]{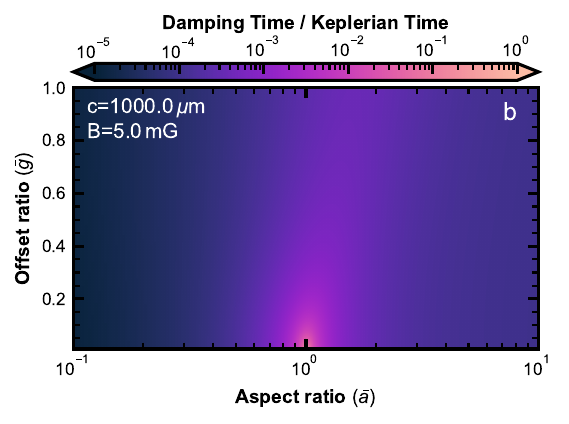} \\
    \includegraphics[width=0.475\textwidth]{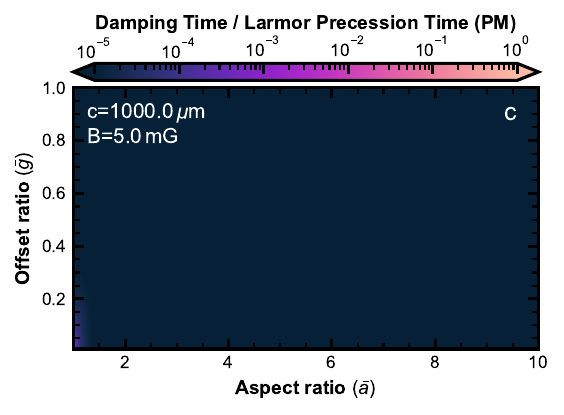} &
    \includegraphics[width=0.475\textwidth]{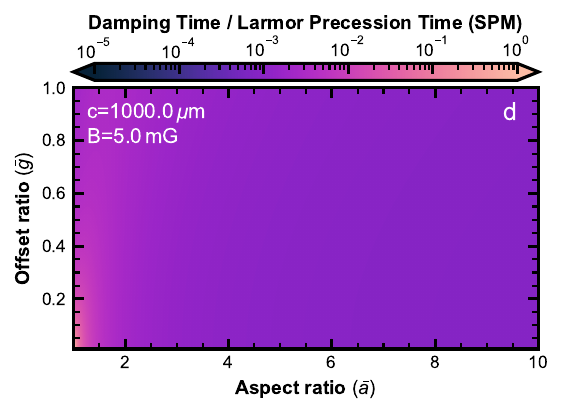}
\end{tabular}
    \caption{The timescale calculations shown in Figure \ref{fig:timescale plots} but using $c=1000\,\mu\text{m}$ and $B=5\,\text{mG}$.}
\end{figure*}

\bibliography{refs}{}
\bibliographystyle{aasjournal}

\end{document}